\documentclass[pra,aps,reprint,superbib,superscriptaddress,twocolumn]{revtex4-2}
\usepackage{amsmath,bm,bbm}
\usepackage{amstext}
\usepackage{epsfig}
\usepackage{xcolor}
\usepackage{subfig}
\usepackage{graphicx}
\usepackage{multirow}
\usepackage{array}
\usepackage{tikz,pgfplots}
\usepackage{MnSymbol}
\usepackage{dsfont}
\usepackage{bbold}
\usepackage{physics}
\usepackage{mathtools}
\usepackage{hyperref}
\usepackage{bm}
\usepackage{xspace}
\usepackage{algorithm2e}
\usepackage{enumitem}
\usepackage{tcolorbox}
\hypersetup{
    colorlinks=true,
    citecolor=gray,
    filecolor=blue,
    linkcolor=blue,
    urlcolor=red
}

\usepackage{mathrsfs}

\definecolor{dgreen}{rgb}{0,.5,0}
\definecolor{dred}{rgb}{.7,.0,.0}

\newcommand{\be}{\begin{eqnarray}}
\newcommand{\ee}{\end{eqnarray}}

\usepackage{xr}

\newcommand{\bmtheta}{{\bm \theta}}

\makeatletter
\renewcommand\@make@capt@title[2]{%
 \@ifx@empty\float@link{\@firstofone}{\expandafter\href\expandafter{\float@link}}%
  {\textbf{#1}}\@caption@fignum@sep#2\quad
}%
\makeatother

\begin{document}

\title{Variational Quantum Subspace Construction via Symmetry-Preserving Cost Functions}
\author{Hamzat A.\ Akande}
\affiliation{Université de Bordeaux, CNRS, LOMA, UMR 5798, F-33400 Talence, France}
\affiliation{ICTP-East African Institute for Fundamental Research, University of Rwanda, Kigali, Rwanda}
\affiliation{ICGM, Univ Montpellier, CNRS, ENSCM, Montpellier, France}
\author{Alexandre Perrin}
\affiliation{Université de Bordeaux, CNRS, LOMA, UMR 5798, F-33400 Talence, France}
\author{Bruno Senjean}
\affiliation{ICGM, Univ Montpellier, CNRS, ENSCM, Montpellier, France}
\author{Matthieu Saubanère}
\email{matthieu.saubanere@cnrs.fr}
\affiliation{Université de Bordeaux, CNRS, LOMA, UMR 5798, F-33400 Talence, France}
\begin{abstract}
Determining low-energy eigenstates in electronic many-body quantum systems is a key challenge in computational chemistry and condensed-matter physics. Hybrid quantum-classical approaches, such as the Variational Quantum Eigensolver and Quantum Subspace Methods, offer practical solutions but face limitations in circuit depth and measurement overhead.
In this article, we propose a variational strategy based on symmetry-preserving cost functions to iteratively construct a reduced subspace for the extraction of low-lying energy states. We show that, under certain conditions, our approach leads to a tridiagonal representation similar to that obtained with the Lanczos algorithm. The iterative process allows control over the trade-off between circuit depth, the number of variational parameters, and the number of measurements required to achieve the desired accuracy, making it suitable for current quantum hardware.
As a proof of concept, we test the proposed algorithms on H$_4$ chain and ring, targeting both the ground-state energy and the charge gap.
 \end{abstract}

 \maketitle

 \section{Introduction}

The many-body problem poses significant challenges in determining electronic low-lying eigenenergies using classical methods. Quantum computing offers a promising alternative to address the exponential complexity~\cite{aspuru-guzikSimulatedQuantumComputation2005, bauerQuantumAlgorithmsQuantum2020, mcardleQuantumComputationalChemistry2020, zimborasMythsQuantumComputation2025}. However, noisy intermediate-scale quantum (NISQ) devices are limited by short qubit coherence times and minimal error correction, making algorithms like phase estimation impractical~\cite{bhartiNoisyIntermediatescaleQuantum2022, omalleyScalableQuantumSimulation2016}. Consequently, hybrid approaches such as the Variational Quantum Eigensolver (VQE) and Quantum Subspace Methods (QSM) have emerged as feasible solutions for current quantum hardware~\cite{peruzzoVariationalEigenvalueSolver2014, mccleanTheoryVariationalHybrid2016,mccleanHybridQuantumclassicalHierarchy2017, stairMultireferenceQuantumKrylov2020, mottaSubspaceMethodsElectronic2024}.

On one hand, the VQE algorithm optimizes a variational trial state encoded in a quantum circuit using an ansatz, broadly classified as physically-motivated or hardware-efficient. Physically-motivated ansätze, such as the unitary coupled-cluster 
(UCC) ansatz~\cite{peruzzoVariationalEigenvalueSolver2014, yungTransistorTrappedionComputers2014, barkoutsosQuantumAlgorithmsElectronic2018, romeroStrategiesQuantumComputing2018, leeGeneralizedUnitaryCoupled2019, evangelistaExactParameterizationFermionic2019, sokolovQuantumOrbitalOptimizedUnitary2020, khamoshiAGPbasedUnitaryCoupled2022, smartQuantumSolverContracted2021, marecatRecursiveRelationsQuantum2023, materiaQuantumInformationDriven2024}, 
are symmetry-preserving (for example
number
of electrons and spin, although the later might still be broken by using trotterized -- also called disentangled -- UCC~\cite{greene2021generalized}) and size-extensive but require deep circuits, which are impractical on current hardware~\cite{ogormanGeneralizedSwapNetworks2019, grimsleyTrotterizedUccsdAnsatz2019, mottaLowRankRepresentations2021}. 
The associated circuit depth can however be reduced using adaptive methods like ADAPT-VQE that selects only impactful excitation 
operators~\cite{grimsleyAdaptiveVariationalAlgorithm2019,liuSimulatingPeriodicSystems2020,tangQubitADAPTVQEAdaptiveAlgorithm2021, feniouOverlapADAPTVQEPracticalQuantum2023, burtonExactElectronicStates2023,mondalOntheflyTailoringRational2023, vaquero-sabaterPhysicallyMotivatedImprovements2024, sunCircuitEfficientQubitExcitationbasedVariational2024, anastasiouTETRISADAPTVQEAdaptiveAlgorithm2024, feniouSparseQuantumState2024, liuPerturbativeVariationalQuantum2024, burtonAccurateGateefficientQuantum2024, ramoaReducingResourcesRequired2024}. 
Despite these improvements, estimating ground-state energies for even small molecules remains challenging due to high gate counts and measurement overhead~\cite{filipReducingUnitaryCoupled2022, grimsleyAdaptiveVariationalAlgorithm2019,vaquero-sabaterPhysicallyMotivatedImprovements2024}.
Shallower circuit can be constructed using Hardware-efficient ansätze (HEA) that employ layers of parametrized single-qubit rotations and entangling gates, enabling practical implementation on NISQ devices~\cite{kandalaHardwareefficientVariationalQuantum2017,ganzhornGateEfficientSimulationMolecular2019}. However, HEA faces scalability challenges due to redundant parameters, lack of symmetry constraints, and barren plateaus, hindering optimization~\cite{mccleanBarrenPlateausQuantum2018, bittelTrainingVariationalQuantum2021, dcunhaChallengesUseQuantum2023, laroccaBarrenPlateausVariational2025,
leonePracticalUsefulnessHardware2024,
perez2024analyzing}. 
To address the aforementioned issues, symmetry-preserving ansätze~\cite{barkoutsosQuantumAlgorithmsElectronic2018, ogormanGeneralizedSwapNetworks2019,burtonAccurateGateefficientQuantum2024, gardEfficientSymmetrypreservingState2020,sekiSymmetryadaptedVariationalQuantum2020, anselmettiLocalExpressiveQuantumnumberpreserving2021, lacroixSymmetryBreakingSymmetry2023,  ruizguzmanRestoringSymmetriesQuantum2024} and advanced optimization strategies~\cite{ryabinkinQubitCoupledCluster2018,
grantInitializationStrategyAddressing2019,
nakanishiSequentialMinimalOptimization2020, sackAvoidingBarrenPlateaus2022, choyMolecularEnergyLandscapes2023,
alvertisClassicalBenchmarksVariational2024,
zambranoAvoidingBarrenPlateaus2024,
novak2025optimization} have been proposed. Additional strategies, such as
adding auxiliary qubits~\cite{yao2025avoiding}, reducing the number of qubits using the quantum matrix product states algorithm~\cite{liu2025matrix},
or using dissipative quantum algorithms~\cite{zapusek2025scaling} have also been developed recently to mitigate the barren plateau issue.
We refer the reader to Refs.~\cite{laroccaBarrenPlateausVariational2025,
cunningham2025investigating} for recent reviews.

On the other hand, QSMs project the Schrödinger equation onto a reduced subspace, combining quantum measurements with classical eigenvalue solvers~\cite{mccleanHybridQuantumclassicalHierarchy2017, yoshiokaGeneralizedQuantumSubspace2022}. The distinction among QSM approaches lies in how the reduced subspace is constructed. For instance, Quantum Subspace Expansion (QSE) constructs subspace vectors using excitation operators, inspired by classical methods like 
Multi-Reference Configuration Interaction Singles and Doubles~\cite{mccleanHybridQuantumclassicalHierarchy2017,takeshitaIncreasingRepresentationAccuracy2020,hugginsNonorthogonalVariationalQuantum2020,urbanekChemistryQuantumComputers2020,yoshiokaGeneralizedQuantumSubspace2022,baekSayNOOptimization2023,umeanoQuantumSubspaceExpansion2025}, but suffers from size-intensivity for large 
systems~\cite{mottaSubspaceMethodsElectronic2024}. Alternatives to the excitation operators include Pauli type
operators~\cite{collessComputationMolecularSpectra2018,bhartiIterativeQuantumassistedEigensolver2021,getelinaQuantumSubspaceExpansion2024} or the use of the  equation-of-motion formalism for excited states~\cite{ollitraultQuantumEquationMotion2020,asthanaQuantumSelfconsistentEquationofmotion2023}.
Quantum Krylov-based methods extend classical eigenvalue solvers to quantum systems, employing Hamiltonian powers to construct subspaces~\cite{parlettSymmetricEigenvalueProblem1998, mottaSubspaceMethodsElectronic2024}. Examples include quantum 
Lanczos~\cite{mcardleVariationalAnsatzbasedQuantum2019,mottaDeterminingEigenstatesThermal2020,yeter-aydenizPracticalQuantumComputation2020}, Chebyshev 
Krylov~\cite{kirbyExactEfficientLanczos2023}, Gaussian power~\cite{zhangMeasurementefficientQuantumKrylov2024}, inverse 
power~\cite{kyriienkoQuantumInverseIteration2020}, Davidson~\cite{tkachenkoQuantumDavidsonAlgorithm2024} and Quantum Filter 
Diagonalization~\cite{parrishQuantumFilterDiagonalization2019,stairMultireferenceQuantumKrylov2020,cohnQuantumFilterDiagonalization2021,cortesQuantumKrylovSubspace2022,epperlyTheoryQuantumSubspace2022,klymkoRealTimeEvolutionUltracompact2022,shenRealTimeKrylovTheory2023,stairStochasticQuantumKrylov2023,kirbyAnalysisQuantumKrylov2024,YoshiokaKrylovdiagonalizationlarge2025}. 
Quantum Krylov based methods require significantly more quantum resources than QSE approaches, due to the construction of the block-encoded unitaries or the need of precise time evolution. However, just like their classical counterpart they excel in approximating precisely dominant eigenvalues with rapid 
convergence~\cite{shenRealTimeKrylovTheory2023, kirbyAnalysisQuantumKrylov2024}. For such approaches, synergy between quantum devices and classical solvers remains central, while challenges like noise and sampling overhead 
persist~\cite{bhartiIterativeQuantumassistedEigensolver2021}.
 
In this contribution, we propose a strategy for designing cost functions that enforce symmetry preservation, regardless of the ansatz used to define the unitary transformation. To obtain a low-energy eigenstate, we introduce an iterative procedure that allows control over the ratio between circuit depth, the number of variational parameters, and the number of measurements required to achieve the desired accuracy. 
In the spirit of QSMs, the iterative process is employed to construct variationally a reduced subspace from which low-lying energy states are extracted. Moreover, we show that the proposed strategy leads to a tridiagonal form analogous to that obtained with the Lanczos algorithm.
As a proof of concept, we test the proposed algorithms on H$_4$ linear and square molecules, targeting both the ground-state energy and the charge gap. Convergence properties in terms of circuit depth and the number of iterations are also analyzed.

 \section{Density-matrix based Variationals principles}
Let us consider the electronic Hamiltonian operator ${\bf H }$ of a system for which the ground state is not degenerate. The Hamiltonian is expanded in the many-body Fock-space basis set $\lbrace |\Phi_i\rangle \rbrace$,
\begin{equation}
{\bf H } = \sum_{ij} H_{ij} |\Phi_i\rangle\langle \Phi_j|,
\end{equation}
where $\lbrace |\Phi_i\rangle \rbrace$ denote Slater determinants or configuration state functions, and we refer to $|\Phi_0\rangle$ as the Hartree--Fock (HF) state.
${\bf  H }$ can be diagonalized using a unitary transformation matrix, ${\bf \overline{H}} = {\bf P^\dagger H P}$, where $\overline{H}_{ij} = E_i \delta_{ij}$ and $E_i$ is the $i^{\rm th}$ eigenvalue sorted in ascending order. 
The eigenvectors are obtained as $|\Psi_i\rangle = {\bf P}|\Phi_i\rangle$,
$|\Psi_0\rangle$ and $E_0$ thus correspond to the ground-state vector and energy, respectively. 
A standard approach to reach the lowest eigenvalue of ${\bf H }$ on a quantum computer relies on the variational principle,
\begin{equation}
\label{eq:varp}
E_0 < E(\bmtheta^*) = \min_{\bmtheta} \langle \Phi_0 | {\bf P }^\dagger(\bmtheta) {\bf H }{\bf P }(\bmtheta) |\Phi_0 \rangle,
\end{equation}
where the unitary transformation ${\bf P }(\bmtheta)$ depends on parameters $\bmtheta$ to be optimized, and $\bmtheta^*$ denotes the optimal parameters.
At the saddle point, $|\Psi_0 (\bmtheta^*)\rangle = {\bf P}(\bmtheta^*)|\Phi_0\rangle$,
and
if ${\bf P}(\bmtheta^*) = {\bf P}$, then $\ket{\Psi(\bmtheta^*)} = \ket{\Psi_0}$ and $E(\bmtheta^*) = E_0$.

\begin{figure}
\resizebox{1.0\columnwidth}{!}{
\includegraphics[scale=1.0]{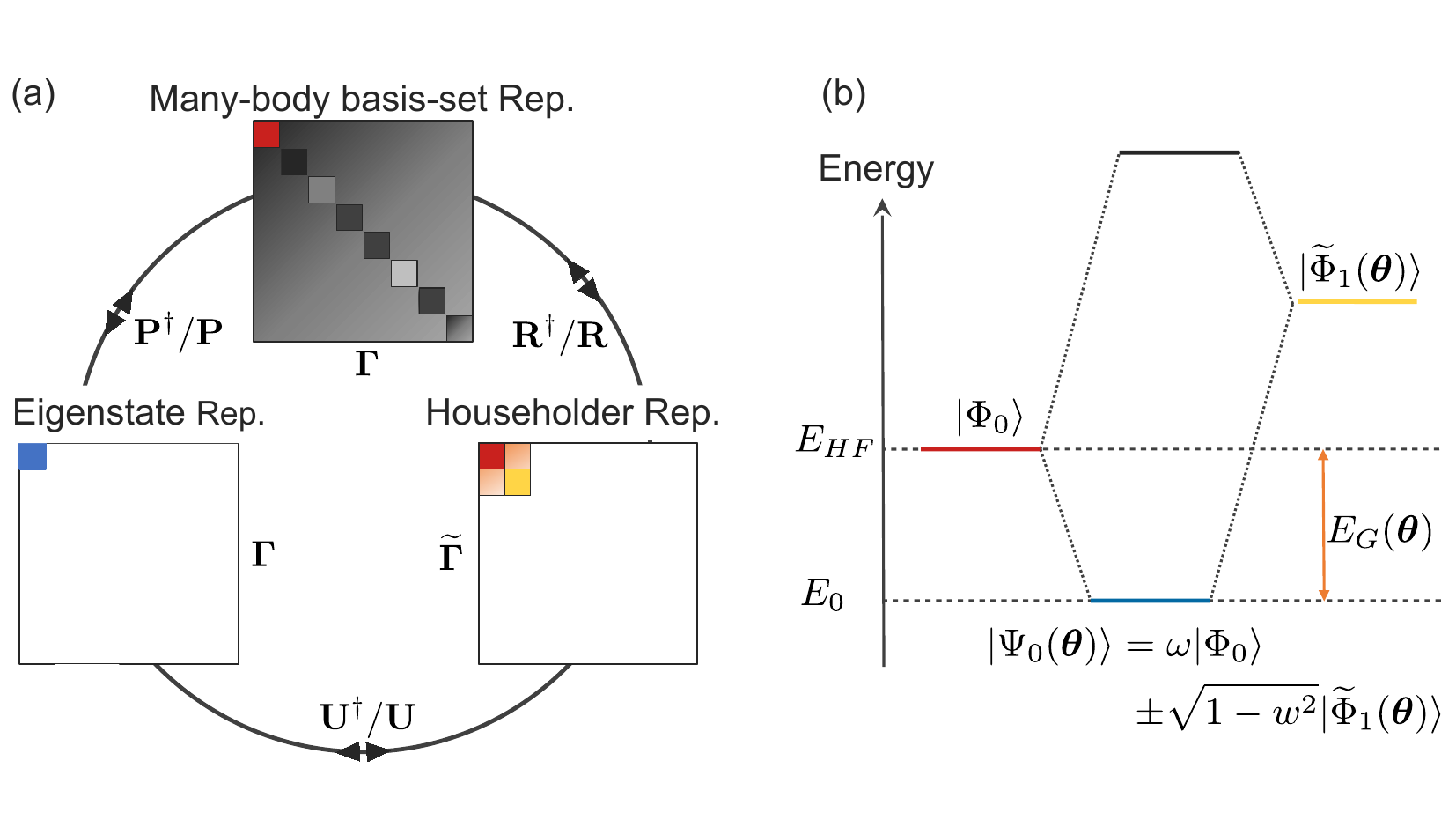}
}
\caption{(a) Schematic representation of the ground-state many-body density matrices ${\bf\Gamma}$, ${\bf \overline{\Gamma}}$, and $\boldsymbol{\widetilde{\Gamma}}$ in the three different representations, namely the many-body basis set, eigenvector, and Householder representation. The unitary transformations ${\bf P}$, ${\bf U}$, and ${\bf R}$ linking the three representations are displayed. (b)  Schematic representation of the two-level decomposition of the ground state. 
We propose to optimize the gain energy $E_G$ coming from the interaction between  the {\it good-guess
  } $|\Phi_0\rangle$ and a variational vector $|\widetilde{\Phi}_1(\bmtheta)\rangle$.    }
\label{fig:fig1}
\end{figure}

%
Alternatively, we rewrite the variational principle in Eq.~(\ref{eq:varp}) in terms of the many-body ground-state density matrices ${\bf \overline{\Gamma}}$ and ${\bf\Gamma}$  of which elements are defined  as
\begin{align}
&  \overline{\Gamma}_{ij} = \langle\Psi_0 |\Psi_i\rangle\langle \Psi_j |\Psi_0\rangle, \quad \Gamma_{ij}= \langle\Psi_0  |\Phi_i\rangle\langle \Phi_j| \Psi_0\rangle.
\end{align}
 As depicted in Fig.~\ref{fig:fig1}(a), $\overline{\Gamma}_{ij} = \delta_{i0}\delta_{j0}$ for all $i$ and $j$. In this framework, the ground-state energy $E_0$ is computed as the following convolution,
\begin{align}
E_0  & = \text{tr}\left({\bf \Gamma}{\bf H} \right) = \text{tr}\left({\bf \overline{\Gamma}}{\bf \overline{H}} \right) = \overline{H}_{00},
\end{align}
and the variational principle in Eq.~(\ref{eq:varp}) corresponds to
\begin{align}
\label{eq:var_pincipleG}
E_0 < E(\bmtheta^*) = \min_{\bmtheta}\left[\text{tr}\left({\bf \overline{\Gamma}} {\bf P}^{\dagger}(\bmtheta) {\bf H} {\bf P}(\bmtheta) \right) \right],
\end{align}
where tr denotes the trace.

We introduce a third intermediate representation in which the first vector $|\Phi_0\rangle$ remains unchanged, and the density matrix
\begin{eqnarray}
\boldsymbol{\widetilde{\Gamma}} =
\begin{bmatrix}
\boldsymbol{\widetilde{\Gamma}}^R & \bf 0 \\
\bf 0 & \bf 0
\end{bmatrix} \label{eq:defR}
\end{eqnarray}
is block-diagonal. $\boldsymbol{\widetilde{\Gamma}}^R$ is a $2\times 2$ matrix that reads
\begin{eqnarray}
  \boldsymbol{\widetilde{\Gamma}}^R = \begin{bmatrix}
    \Gamma_{00} & \widetilde{\Gamma}_{10}   \\
    \widetilde{\Gamma}_{01}  &  \widetilde{\Gamma}_{11}
  \end{bmatrix} = \begin{bmatrix}
    \omega^2  & \sqrt{\omega^2(1-\omega^2)} \\
    \sqrt{\omega^2(1-\omega^2)} & 1-\omega^2
  \end{bmatrix}   ,     \label{eq:defGD}
\end{eqnarray}
where $\omega^2 = |\langle\Psi_0|\Phi_0\rangle|^2$ represents the weight of $|\Phi_0\rangle$ in the ground state. 
This representation is called the Householder representation since, among all unitary transformations ${\bf R}$ that provide the form of $\widetilde{\boldsymbol{\Gamma}}$ in Eq.~(\ref{eq:defR}),
\begin{equation}
\boldsymbol{\widetilde{\Gamma}} = {\bf R} \boldsymbol{\Gamma} {\bf R},
\end{equation}
with $|\Phi_0\rangle = {\bf R} |\Phi_0\rangle$, the Householder reflection (${\bf R} = {\bf R}^\dagger$) is uniquely defined from the first column of $\boldsymbol{\Gamma}$. Note that this transformation has been recently used in the context of embedding methods in electronic structure theory~\cite{sekaranHouseholderTransformedDensity2021a,yalouzQuantumEmbeddingMultiorbital2022,marecatUnitaryTransformationsDensity2023,marecatVersatileUnitaryTransformation2023}.
Using this transformation, the ground-state vector is decomposed as
\begin{equation}
  \label{eq:GSvect}
  |\Psi_0\rangle = \omega |\Phi_0\rangle + \nu \sqrt{1 - \omega^2}{\bf R} |\Phi_1\rangle, 
\end{equation}
where $\nu = \pm 1$ is a relative phase factor,
and the ground-state energy reads, in this new representation,
\begin{align}
E_0 & = \text{tr}\left(\boldsymbol{\widetilde{\Gamma}}\widetilde{\bf H} \right)= \text{tr}\left(\boldsymbol{\widetilde{\Gamma}}^R\widetilde{\bf H}^R \right),
\end{align}
where $\widetilde{\bf H} = {\bf R H R}$ and
\begin{eqnarray}
  \widetilde{\bf H}^R = \begin{bmatrix}
    H_{00} & \widetilde{H}_{10}   \\
    \widetilde{H}_{01}  &  \widetilde{H}_{11}
  \end{bmatrix}.
\end{eqnarray}
It follows that the variational principle in Eq.~(\ref{eq:var_pincipleG}) can be rewritten in the Householder representation as
\begin{align}
E_0 < E(\bmtheta^*, \omega^*) = \min_{\omega,\bmtheta} \left[\text{tr}\left(\boldsymbol{\widetilde{\Gamma}}^R(\omega) \widetilde{{\bf H}}^R(\bmtheta) \right) \right],
\label{eq:var_DM_Hous}
\end{align}
where 
$\bmtheta^*$ and $\omega^*$ are the minimizing parameters, and
\begin{eqnarray}
  \widetilde{ {\bf H}}^R(\bmtheta)= \begin{bmatrix}
    H_{00} & \widetilde{H}_{10}(\bmtheta)   \\
    \widetilde{H}_{01}(\bmtheta)  & \widetilde{H}_{11}(\bmtheta)
  \end{bmatrix} \label{eq:defHtheta1},
\end{eqnarray}
with $\widetilde{\mathbf{H}}(\bmtheta) = {\bf R}^\dagger(\bmtheta)\mathbf{H}{\bf R}(\bmtheta)$.
By introducing the Householder representation, we have split the variational process into two sub-processes: 1) the optimization of the reflection ${\bf R}(\bmtheta)$ to block-diagonalize ${\bm \Gamma}$ into $\widetilde{\bf \Gamma}$, and 2) the optimization of the unitary ${\bf U}(\omega)$ to diagonalize ${\widetilde{\bf H}^R} (\bmtheta)$, i.e. ${\bf P} = {\bf RU}$, see Fig.~(\ref{fig:fig1}). Note that since $\widetilde{\bf \Gamma}$ is idempotent, ${\bf U}(\omega)$ is only defined by a single parameter $\omega$. 
Optimizing ${\bf R}(\bmtheta)$ consists in finding the complementary vector $|\widetilde{\Phi}_1(\bmtheta)\rangle = {\bf R}(\bmtheta)|\Phi_1 \rangle$ of $|\Phi_0\rangle$ that composes the ground state, as shown in Eq.~(\ref{eq:GSvect}).

\section{Iterative and variational symmetry-preserving quantum eigensolver}
\subsection{Symmetry-preserving cost functions}
\label{subsec:Cost}
Let us make the assumption that $|\Phi_0\rangle$ is a binding state i.e. $H_{00} = \langle \Phi_0 |{\bf H} |\Phi_0\rangle \leq 0$ and additionally that it is a {\it good-guess}, i.e. that $|\Phi_0\rangle$ as an overlap $\omega^2 = |\langle \Phi_0|\Psi_0\rangle|^2 \geq 0.5$ with the ground state.
In that case, instead of using the energy as a cost function to be minimized, we consider the gain energy,
\begin{eqnarray}
E_G(\bmtheta) = \frac{\Delta(\bmtheta)}{2} \left[ 1 - \sqrt{1 + \frac{4\widetilde{H}_{01}(\bmtheta)\widetilde{H}_{10}(\bmtheta)}{\Delta(\bmtheta)^2}} \right],
\label{eq:Egain}
\end{eqnarray}
where $\Delta(\bmtheta) = \widetilde{H}_{11}(\bmtheta) - H_{00}$. $E_G(\bmtheta)$ is negative if $\Delta(\bmtheta) > 0$ and that is computed in the two-level system composed of $|\widetilde{\Phi}_1(\bmtheta)\rangle$ and $|\Phi_0\rangle$, as shown in Fig.~\ref{fig:fig1}(b).
It corresponds to the energy difference between the eigenstate of the two-level system $|\Psi_0\rangle$ and $|\Phi_0\rangle$, i.e. $E_G(\bmtheta) = \langle \Psi_0(\bmtheta)| {\bf H}| \Psi_0(\bmtheta)\rangle - \langle \Phi_0| {\bf H}| \Phi_0\rangle$. If  $|\Phi_0\rangle$ is a good guess, the ground-state energy $E_0 = \langle \Psi_0| {\bf H}| \Psi_0\rangle$ is related with the Hartree--Fock energy $H_{00}$ and the gain energy as
\begin{equation}
  \label{eq:E0}
E_0(\bmtheta^*) = H_{00} + E_G(\bmtheta^*),
\end{equation}
for the optimal parameter $\bmtheta^*$. 
From a numerical point of view, using $E_G(\bmtheta)$ as a cost function for optimization may lead to instability when $\Delta(\bmtheta) \to 0$, i.e., 
when the spectrum exhibits degeneracy or when the target state is not the lowest eigenvalue of the spectrum but the lowest-eigenvalue of a specific symmetry-restricted subspace. Alternatively, the interaction energy 
\begin{eqnarray}
E_I(\bmtheta) = - \sqrt{ \widetilde{H}_{01}(\bmtheta)\widetilde{H}_{10}(\bmtheta)}
\label{eq:Eint}
\end{eqnarray}
can be also considered as a relevant cost function.
In particular, for large $|\Delta(\bmtheta)|$ compared to $|\widetilde{H}_{10}(\bmtheta)|$, both cost functions are expected to lead to the same saddle point as  
\begin{equation}
\frac{\partial E_G(\bmtheta)}{\partial \bmtheta} \propto \frac{\partial E_I(\bmtheta)}{\partial \bmtheta}. \label{eq:cost_sim}
\end{equation}
Moreover, we show in the following that minimizing $E_I(\bmtheta)$ with the condition that  
$\langle \Phi_0 | {\bf R}(\bmtheta) \Phi_0 \rangle = 1$  
corresponds to emptying the first column/row of the Hamiltonian up to the diagonal term and the first off-diagonal element i.e.,  
$\widetilde{H}_{0j}(\bmtheta) = \sum_{kl} R_{0k}(\bmtheta) H_{kl}R_{lj}(\bmtheta) = 0, \quad \forall j > 1$, similarly as one obtains by using the corresponding Householder Reflection.\\
{\it Proof:}  
Let us consider the first diagonal element of the square of the Hamiltonian in the Householder representation $\boldsymbol{ \widetilde{H}}$ and in the variational representations $\boldsymbol{ \widetilde{H}}(\bmtheta) $,
\begin{eqnarray}
  [\boldsymbol{\widetilde{H}}^2]_{00} &= &H_{00}^2 + \widetilde{H}_{01}\widetilde{H}_{10}, \\
  {[\boldsymbol{\widetilde{H}}(\bmtheta)^2]}_{00} &= &\widetilde{H}_{00}(\bmtheta)^2 + \widetilde{H}_{01}(\bmtheta)\widetilde{H}_{10}(\bmtheta)  \nonumber\\
 &  & + \sum_{j > 1} \widetilde{H}_{0j}(\bmtheta)\widetilde{H}_{j0}(\bmtheta).
\end{eqnarray}
Because the condition $\langle \Phi_0 | {\bf R} (\bmtheta)|  \Phi_0 \rangle= 1$ imposes that $\widetilde{H}_{00} = H_{00}$ and $[\boldsymbol{\widetilde{H}}^2]_{00} = [\boldsymbol{\widetilde{H}}(\bmtheta)^2]_{00}$, it follows that 
\begin{equation}
  \label{eq:tridiag}
\widetilde{H}_{01}\widetilde{H}_{10} > \widetilde{H}_{01}(\bmtheta)\widetilde{H}_{10}(\bmtheta),
\end{equation}
except if for all $j >1$, $\widetilde{H}_{j0}(\bmtheta) = 0$. In that latter case, $\widetilde{H}_{01}\widetilde{H}_{10} = \widetilde{H}_{01}(\bmtheta)\widetilde{H}_{10}(\bmtheta)$. Consequently, the saddlepoint of minimizing $E_I(\bmtheta)$ corresponds to the case where for all $j>1$,  $\widetilde{H}_{j0}(\bmtheta^*) = \widetilde{H}_{j0} = 0$ \cite{fnote}.  

A first important remark deals with the fact that cost functions,  
$C (\bmtheta) = E_G(\bmtheta)$ or $E_I(\bmtheta)$, cancel out if  
$|\widetilde{\Phi}_1(\bmtheta)\rangle$ and $|\Phi_0\rangle$ do not belong to the same symmetry-defined subspace of the Hamiltonian, i.e., if  
\[
\widetilde{H}_{01} (\bmtheta) = \langle \Phi_0 |{\bf H} | \widetilde{\Phi}_1(\bmtheta)\rangle = 0.
\]  
In this way, the desired symmetries are imposed by construction in the guess vector $|\Phi_0\rangle$. This property is expected to drastically simplify the energy landscape since only the states belonging to the same symmetry-subspace as $|\Phi_0\rangle$  can contribute to $C(\bmtheta)$ while states that does not belong to the targeted symmetry subspace have a zero contribution. Consequently, it is naturally expected that for the optimal parameters $\bmtheta^*$ minimizing $C$, the state  
$|\widetilde{\Phi}_1(\bmtheta^*)\rangle$ exhibits the same quantum numbers (e.g., $\langle N \rangle$, $\langle S^2 \rangle$, $\langle S_z \rangle$) as $|\Phi_0\rangle$, regardless of the ansatz used to construct ${\bf R}(\bmtheta)$. The use of symmetry-preserving cost functions $C (\bmtheta)$ is therefore highly compatible with HEA.  

A second important remark relates to the fact that, in contrast to the VQE algorithm which relies on the minimization of energy, minimizing both $C(\bmtheta)$ does not guarantee to reach the ground state. 
Indeed, $E_G(\bmtheta)$ remains negative only for positive $\Delta(\bmtheta)$. Consequently, minimizing $E_G(\bmtheta)$ decreases the energy but constrains  
$
\omega(\bmtheta) = |\langle \Phi_0 | \Psi_0 (\bmtheta) \rangle|^2
$ 
to be larger than $0.5$ to ensure that $\Delta(\bmtheta)$ is positive. It follows that if $|\Phi_0 \rangle$ is not a {\it good guess}, minimizing $E_G(\bmtheta)$ does not lead to the ground state but rather to the lowest-energy state within the symmetry-subspace where $\omega(\bmtheta) > 0.5$.  
Similarly, using $E_I(\bmtheta)$ as a cost function does not guarantee convergence to the ground state but instead promotes maximal interaction between $|\Phi_0 \rangle$ and $|\widetilde{\Phi}_1(\bmtheta) \rangle$ within the same symmetry-subspace.  
To overcome this limitation, an iterative scheme is proposed in the following.

\subsection{Iterative and variational quantum eigensolver (IVQE)}
\label{sec:IVQE}

The {\it good-guess} condition is replaced by the less constraining requirement for $|\Phi_0\rangle$ to be a binding state that has a non-zero overlap with the targeted ground state. To obtain the ground state, we propose an iterative loop condition compatible with the cost function $C(\bmtheta)$. More precisely, we set up an iterative process such that the ground state $|\Psi_0^{(n)}(\bmtheta^{(n)*})\rangle$ of the two-level system (see Fig.~\ref{fig:fig1}(b)) obtained at the end of an iteration $n$ is used as the \textit{guess} vector $|\Phi_0^{(n+1)}\rangle$ for the next iteration $n+1$. The algorithm is initialized at iteration $n = 0$ with $|\Psi_0^{(n=0)} \rangle = \ket{\Phi_0}$. The process for $n > 0$ can be decomposed into the following steps:
\begin{enumerate}
\item {\it Loop condition and guess vector definition:} 
$|\Psi_0^{(n-1)} \rangle \to |\Phi_0^{(n)}\rangle$ becomes the new \textit{guess} vector for iteration $n$.
\item {\it Find optimal parameter $\bmtheta^{(n)*}$ for the unitary} ${\bf R}(\bmtheta^{(n)})$.
  We optimize the cost function $C(\bmtheta^{(n)})$.
  The classical optimization algorithm requires multiple evaluations of the cost function.\\
  Inner loop for cost function computation:
  \begin{enumerate}
  \item {\it Construct a quantum circuit} (quantum computer) representing the variational trial state $|\widetilde{\Phi}_1^{(n)}(\bmtheta^{(n)})\rangle$.
  \item {\it Measure} (quantum computer) all the necessary elements of the overlap and Hamiltonian matrix 
    and the Hamiltonian:
    \begin{align}
      &\mathcal{S}_{pn}=\langle  \widetilde{\Phi}_1^{(p)}(\bmtheta^{(p)*})|\widetilde{\Phi}_1^{(n)}(\bmtheta^{(n)})\rangle,\\
        & \nonumber\\
      &\mathcal{H}_{pn}= \langle \widetilde{\Phi}_1^{(p)}(\bmtheta^{(p)*})|\mathbf{H} | \widetilde{\Phi}_1^{(n)}(\bmtheta^{(n)})\rangle \label{eq:mathcalH}
    \end{align}
      for $ 0\leq p \leq n $ and considering $\vert \widetilde{\Phi}_1^{(0)} (\bmtheta^{(0)})\rangle = \ket{\Phi_0}$.
    \item {\it Orthogonalize} (classical computer)
      the  variational trial state $|\widetilde{\Phi}_1^{(n)}\rangle$ with respect to the overlap matrix $\boldsymbol{\mathcal{S}}$. Update all elements of $\boldsymbol{\mathcal{H}}$ in the orthonormalized basis set.
  \item {\it Compute the cost function} (classical computer) by constructing the matrix ${\widetilde{\bf H}^R} (\bmtheta^{(n)})$ from the elements of $\boldsymbol{\mathcal{H}}$ in Eq.~(\ref{eq:mathcalH}). Then, use Eq.~(\ref{eq:Egain}) or Eq.~(\ref{eq:Eint}) to obtain the cost function $C(\bmtheta^{(n)})$. \label{algo:C}
  \end{enumerate}
The minimizing parameters are denoted by $\bmtheta^{(n)*}$.

\item {\it Two-level system ground state.} (classical computer) \label{algo:twolevel}
  We diagonalize ${\widetilde{\bf H}^R} (\bmtheta^{(n)*})$ in the subspace spanned by $|\Phi_0^{(n)}\rangle (=| \Psi_0^{(n-1)}\rangle)$ and $|\widetilde{\Phi}_1^{(n)}(\bmtheta^{(n)*})\rangle$, obtaining the optimal $\omega^{(n)*}$ [see Eq.~(\ref{eq:GSvect})]. This leads to:
  \begin{align}
    |\Psi_0^{(n)}\rangle = & \omega^{(n)*} |\Psi_0^{(n-1)}\rangle \nonumber \\
    & \pm \sqrt{1 - {\left(\omega^{(n)*}\right)}^2} | \widetilde{\Phi}_1^{(n)}(\bmtheta^{(n)*})\rangle.
\end{align}
Ground state properties at iteration $n$  such as the energy can be evaluated by using the diagonal form of ${\widetilde{\bf H}^R} (\bmtheta^{(n)*})$
\[
  \label{eq:E0n}
E_0^{(n)} = \langle \Psi_0^{(n)} | {\widetilde{\bf H}^R} (\bmtheta^{(n)*})|\Psi_0^{(n)}\rangle.
\]
\item {\it Check convergence.} (classical computer) 
If ${\left(\omega^{(n)*}\right)}^2 > 1 - \varepsilon_w$ and/or the minimized cost function $C(\bmtheta^{(n)*}) > -\varepsilon_e$, the process is considered converged. Here, $\varepsilon_w$ and $\varepsilon_e$ are positive convergence criterion parameters that must be user-defined. Otherwise, the algorithm restarts at step 1 with $n = n+1$.
\end{enumerate}

\begin{figure}
  \centering
  \resizebox{0.75\columnwidth}{!}{\includegraphics[scale=1]{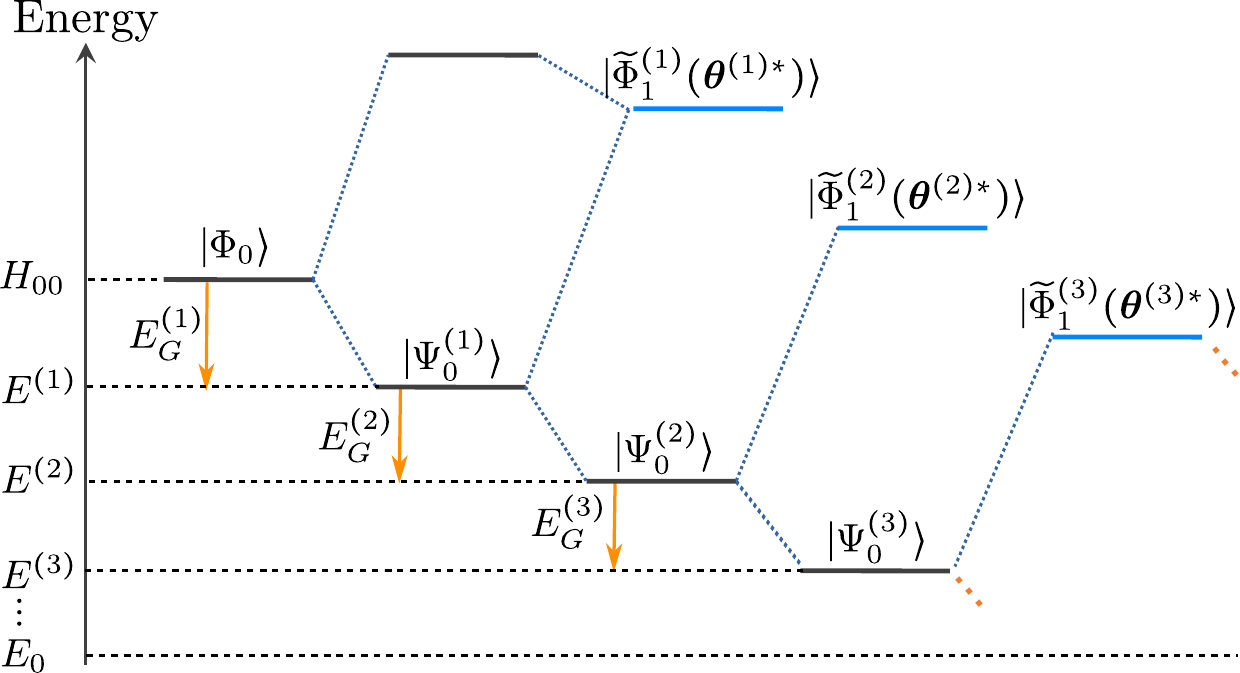}}\\
  \caption{Schematic depiction of the iterative algorithm (top panel). Only the states $\lbrace |\widetilde{\Phi}_1^{(n)}(\bmtheta^{(n)})\rangle \rbrace$ in blue are optimized on a quantum computer to minimize the cost function $C(\bmtheta^{(n)})$.}
  \label{fig:fig_iterative}
\end{figure}

This procedure is depicted in Fig.~\ref{fig:fig_iterative}. The convergence criteria are based on the fact that if $\omega^{(n)*} \neq 1$ or if the cost function is strictly negative $C(\bmtheta^{(n)*}) \neq 0$, then $|\Psi_0^{(n-1)}\rangle$ is not an eigenstate. On the contrary, if $\omega^{(n)*} = 1$ or $C(\bmtheta^{(n)*}) = 0$, then $|\Psi_0^{(n-1)}\rangle$ is an eigenstate in the same symmetry subspace as the initial guess $\ket{\Phi_0}$. Note that only $|\widetilde{\Phi}_1^{(n)}(\bmtheta^{(n)})\rangle$ are constructed on the QC, while $|\Psi_0^{(n)}\rangle$ are never constructed explicitly but retrieved using ${\widetilde{\bf H}^R} (\bmtheta^{(n)*})$.

This algorithm will be referred to as the Iterative Variational Quantum Eigensolver (IVQE) algorithm. It falls into the class of VQEs, as each iteration is a new variational process with an updated \textit{guess}-vector, for which the cost function $C(\bmtheta^{(n)})$  is minimized. 
Both the gain and interaction energy can be used as the cost function. The former is expected to speed up the convergence, while the latter is expected to provide more stable numerical optimization. 

\subsection{ Variational quantum subspace method (VQSM)}

We can exploit the fact that at each iteration $n$, the set of optimal trial states $\left\{ \vert \widetilde{\Phi}_1^{(p)*}(\bmtheta^{(p)*}) \rangle\right\}$ optimized iteratively in  $0 \leq p \leq n$  iterations, constitutes a reduced orthonormal 
basis in which the Hamiltonian is expanded. It follows that we can revise steps \ref{algo:C} and \ref{algo:twolevel} of the algorithm presented in the previous section~\ref{sec:IVQE} in the spirit of quantum 
subspace method. More precisely, instead of using the two-level system Hamiltonian  $\widetilde{\mathbf{H}}^{R(n)}(\bmtheta^{(n)})$  at iteration $n$ to compute the cost function and define the iteration ground state, we could 
diagonalize the $(n + 1) \times (n + 1)$ Hamiltonian  matrix $\boldsymbol{\mathcal{H}}^{(n)}$ within the subspace $\left\{ \vert \widetilde{\Phi}_1^{(p)}(\bmtheta^{(p)}) \rangle \right\}$ . In parallel, we also need to adapt the cost functions, Eq.~(\ref{eq:Egain}) or Eq.~(\ref{eq:Eint}), to the subspace expansion approach. 
The cost function Eq.~(\ref{eq:Egain}) is 
not changed and consists of the energy gain computed in the two-level system constituted from the ground state $|\Psi_0^{(n-1)} \rangle$ of $\boldsymbol{\mathcal{H}}^{(n-1)}$ 
of the previous iteration and the variational trial vector $| \widetilde{\Phi}_1^{(n)}(\bmtheta^{(n)})\rangle$. It takes 
the same form as Eq.~(\ref{eq:Egain}) but $\widetilde{H}_{01}(\bmtheta) = \langle  \widetilde{\Phi}_1^{(n)}(\bmtheta^{(n)}) |{\bf H} | \Psi_0^{(n-1)} \rangle$ and  $\widetilde{H}_{11}(\bmtheta) = \langle  \widetilde{\Phi}_1^{(n)}(\bmtheta^{(n)}) |{\bf H} |  \widetilde{\Phi}_1^{(n)}(\bmtheta^{(n)}) \rangle$.
In the same line, Eq.~(\ref{eq:Eint}) can also be used to maximize the interaction energy with the ground state of the 
previous iteration using $\widetilde{H}_{01}(\bmtheta) = \langle  \widetilde{\Phi}_1^{(n)}(\bmtheta^{(n)}) |{\bf H} | \Psi_0^{(n-1)} \rangle$. 

Interestingly, following property~(\ref{eq:tridiag}), we
can define another type of cost function,
\begin{equation}
  \label{eq:cost_Tri}
E_I'(\bmtheta^{(n)}) = -\sqrt{ \mathcal{H}_{n(n-1)}^{(n)}(\bmtheta) \mathcal{H}_{(n-1)n}^{(n)}(\bmtheta)}
\end{equation}
which upon minimization provides a tridiagonal form of $\boldsymbol{\mathcal{H}}$. Ultimately, given a fixed starting vector, it leads to a tridiagonalization of the Hamiltonian equivalent to the Householder, Givens, and Lanczos tridiagonalization methods on classical computers~\cite{wilkinsonAlgebraicEigenvalueProblem1992, golubMatrixComputations1996, parlettSymmetricEigenvalueProblem1998, golubEigenvalueComputation20th2000}, since Pick and Tomasek showed that given a starting vector the tridiagonal form is unique~\cite{pickEquivalenceThreeApproaches1982}. 
Consequently, the convergence performances should also be comparable.
Consequently, we can expect from this analysis that the variational iterative algorithm we propose will converge rapidly to the dominant eigenvalue, similar to classical Lanczos/Householder tridiagonalization scheme. 

Altogether, it follows that steps \ref{algo:C} and \ref{algo:twolevel} of the IVQE algorithm can be replaced by:
\begin{enumerate}
  \item [2.] {\it \dots}
  \begin{enumerate}
  \item[(d')]{\it Compute the cost function:} (classical computer) First diagonalize $\boldsymbol{\mathcal{H}}^{(n)}$, then use the generalization of Eq.~(\ref{eq:Egain}) or Eq.~(\ref{eq:Eint}) to obtain the cost function $C(\bmtheta^{(n)})$.
   \end{enumerate}
   \item[3'.] {\it Subspace expansion ground state:} (classical computer)
     Once the parameters $\bmtheta^{(n)*}$ optimized, we diagonalize $\boldsymbol{\mathcal{H}}^{(n)}$ expanded in the orthonormal subspace spanned by $\left \lbrace |\widetilde{\Phi}_1^{(p)}(\bmtheta^{(p)*})\rangle \right \rbrace$, $0 \leq p \leq n$. The ground state of $\boldsymbol{\mathcal{H}}^{(n)}$ is denoted $|\Psi_0^{(n)}\rangle$. We compute $\omega^{(n)*} = \langle\Psi_0^{(n-1)} |\Psi_0^{(n)} \rangle$ as the fidelity of the obtained ground state with respect to the ground state obtained at the previous iteration. Also the ground-state properties at iteration $n$ such as the energy can be evaluated as
     \[
      \label{eq:E0n2}
E_0^{(n)} = \langle \Psi_0^{(n)}|\boldsymbol{\mathcal{H}}^{(n)}|\Psi_0^{(n)}\rangle.
\]
\end{enumerate}

This algorithm will be denoted as the Variational Quantum Subspace Method (VQSM) algorithm. It falls in the class of quantum subspace methods, as in each iteration the subspace is enlarged, optimizing gain or interaction energy with the ground state of the previous iteration.
In terms of performance, VQSM is expected to converge faster to the dominant eigenvalue than IVQE, at the costs of an additional classical diagonalization of the subspace Hamiltonian.
Regarding the classical computational overhead, we acknowledge that, unlike IVQE, VQSM requires a matrix diagonalization step at each iteration. However, this overhead remains moderate in practice, as the size of the matrix to be diagonalized is determined by the number of iterations, which typically remains small due to the geometric convergence behavior of such methods. Importantly, this subspace dimension usually grows only polynomially with system size. Thus, while VQSM introduces more classical post-processing, it compensates by offering significantly improved accuracy and convergence.
In this sense, VQSM sits at the intersection of VQE and quantum subspace methods, combining the advantages of both approaches: it leverages very shallow quantum circuits enabled by symmetry-preserving cost functions typical of HEA-based VQE, while also inheriting the robustness, noise resilience, and fast convergence associated with QSM. For large-scale applications, a hybrid strategy combining IVQE in the early iterations and switching to VQSM in later stages could effectively balance classical overhead and algorithmic accuracy.

\subsection{Practical implementation and measurements}

\begin{figure}
  \centering
  \resizebox{0.8\columnwidth}{!}{\includegraphics[scale=1.]{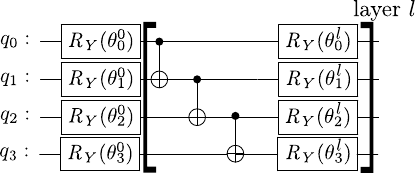}}
  \caption{Schematic representation of the HEA used in this work. The circuit inside the square brackets representing one layer $l$ is repeated $N_L$ times.}
  \label{fig:fig_HEA}
\end{figure}

\textit{Construction of the Trial Variational State on the Quantum Computer.}
Ideally, the trial state $|\widetilde{\Phi}_1(\bm{\theta})\rangle$ is prepared by applying a reflection ${\bf R}(\bm{\theta})$ to an ``easy-to-prepare'' initial state, such as the Hartree--Fock (HF) 
state -- similar to the procedure used in standard Householder tridiagonalization.
Using $E_I$ is expected to mitigate optimization difficulties since it is linear in ${\bf R}$. Indeed, since ${\bf R}$ can be defined up to a phase factor, the quantity $\widetilde{H}_{01}({\bf R}) = \langle \Phi_0 | {\bf H R} | \Phi_1 \rangle$, where $|\Phi_0\rangle$ and $|\Phi_1\rangle$ are fixed vectors, can be arbitrarily chosen to be positive.
By contrast, $E_G$ is generally non-convex (quadratic in ${\bf R}$), although it tends toward convexity close to the non-interacting limit, suggesting that severe optimization challenges may be avoided in that regime.
However, despite recent progress, implementing a reflection on a quantum computer remains technically challenging, as it typically requires multi-controlled NOT gates~\cite{ortsStudyingCostNqubit2022, claudonPolylogarithmicdepthControlledNOTGates2024}.
That said, Householder reflections is unique among many other unitary transformations capable of block-diagonalizing ${\bf H}$ while preserving $|\Phi_0\rangle$. Therefore, rather than constraining the transformation to be a reflection, we construct the trial state $|\widetilde{\Phi}_1(\bm{\theta})\rangle$ using a variational unitary quantum circuit applied to a simple reference state.
Optimally, the variational circuit would exactly reproduce the effect of the targeted reflection. 
While the cost functions $E_I$ ($E_G$) introduced above shows convex properties (or nearly convex properties), respectively, the discrepancy between the actual circuit and the ideal Householder reflection likely results in a highly non-convex function, thus complicating the optimization.
As a consequence, replacing the explicit construction of ${\bf R}(\bm{\theta})$ with a variational unitary circuit may introduce the typical issues encountered in variational methods -- such as barren plateaus and the proliferation of local minima -- well documented in other ansatz-based approaches~\cite{laroccaBarrenPlateausVariational2025}.

{\it Hardware efficient Ansatz.}
The quantum circuit in Fig.~\ref{fig:fig_HEA} aims to construct $|\widetilde{\Phi}_1(\bm{\theta})\rangle = {\bf U}_{qc}(\bmtheta)|\Phi_I\rangle$
 and is constituted by an initialization, i.e the `easy' construction of $|\Phi_I\rangle$ followed by the unitary circuit ${\bf U}_{qc}(\bmtheta)$. 
For the initialization step, one might choose an excited Slater determinant $|\Phi_I\rangle$ that has an overlap with the ground state, i.e. that belong to the same symmetry subspace than $|\Phi_0\rangle$.
However to remain general, here we just initialize the circuit with the zero state. 
To construct the unitary part ${\bf U}_{qc}(\bmtheta)$, we exploit the fact that the cost functions $E_G(\bmtheta)$ and $E_I(\bmtheta)$ enforce $|\widetilde{\Phi}_1(\bmtheta)\rangle$  to be in the same symmetry-subspace as the initial {\it guess} vector $|\Phi_0\rangle$. 
This allows in principle the use of any ansatz, including HEA to reduce circuit depth. 
As a sake of simplicity and controllability, we construct ${\bf U}_{qc}(\bmtheta)= {\rm HEA}$ as a  first layer of $R_y(\theta)$ gates on each qubit  followed by  series of $N_L$ layers, each containing two-qubit entangling gates and single-qubit rotational gates. Note if the circuit is initialized with an excited state, the first $R_y(\theta)$ layer becomes unnecessary.
For instance, a simple case depicted in Fig.~\ref{fig:fig_HEA} consists in the 1D R$_y$CNOT HEA~\cite{kandalaHardwareefficientVariationalQuantum2017} that we propose to use in the following as a proof of concept.  
The ansatz consists in  $N_L$ layers containing an entangling operator. The entangling operator consists of a linear chain of $N_{\text{ent}}=(N_{\rm qubit}-1)$ two-qubit CNOT gates, employing a linear entanglement strategy. Each layer also includes a set of $R_y(\theta)$ rotation gates applied to all qubits. The total number of variational parameters is $(N_L+1)N_{\rm qubit}$, while the number of CNOT gates is $N_L (N_{\rm qubit}-1)$. Since qubit connectivity affects the number of entangling gates per layer, $N_{\text{ent}}$ can be adjusted accordingly, becoming $N_{\text{ent}} = N_{\rm qubit} (N_{\rm qubit}-1)/2$ for a fully connected setup.

{\it Measurement.}
Using $E_G(\bm\theta)$, $E_I(\bm\theta)$, or $E_I'(\bm\theta)$ as the objective function requires additional measurements compared to standard VQE. Specifically, it necessitates measuring the elements of $\boldsymbol{\mathcal{H}}$ and $\boldsymbol{\mathcal{S}}$. The diagonal elements can be obtained through standard projective measurements.
The measurement of off-diagonal matrix elements has received particular attention in the context of Quantum Subspace Methods~\cite{mottaSubspaceMethodsElectronic2024}, as these quantities are essential for constructing effective Hamiltonians in a reduced basis. A standard approach to estimating such matrix elements of the form $O_{\alpha\beta} = \langle v_0 | \hat{U}_\alpha^\dagger \hat{O} \hat{U}_\beta | v_0 \rangle$, where $\hat{U}_\alpha$, $\hat{U}_\beta$ are unitary operators and $\hat{O}$ is typically the Hamiltonian or the identity, is the Hadamard test. This method requires an ancilla qubit and the implementation of controlled versions of $\hat{U}_\alpha$ and $\hat{U}_\beta$, which significantly increases the circuit depth and the number of entangling gates. Specifically, each single-qubit gate in $\hat{U}$ must be replaced by its controlled counterpart (requiring two CNOT gates), and each CNOT gate must be promoted to a Toffoli gate, which typically decomposes into six single-qubit gates and four or six CNOT gates.
To mitigate these costs, recent alternatives have been developed that avoid the Hadamard test. Notably, Cortés and Gray~\cite{cortesQuantumKrylovSubspace2022} proposed a measurement-based protocol that reconstructs the off-diagonal matrix element $O_{\alpha\beta}$ from fidelity measurements, using overlaps between specially prepared quantum states. Alternatively, simplifications are possible by exploiting physical symmetries. For instance, if spin conservation holds, one can avoid implementing costly controlled time-evolution operators by instead performing a controlled initialization of the reference state~\cite{YoshiokaKrylovdiagonalizationlarge2025}.
Consequently, while the use of the Hadamard test is expected to significantly increase the circuit depth and CNOT count, these alternative strategies allow for substantial resource savings and make the method more suitable for near-term applications.

 {\it Circuit depth.}
  To contextualize the performance of our approach, we compare the CNOT gate counts required to achieve chemical accuracy with those reported for state-of-the-art adaptive methods such as ADAPT-VQE. For example, the original fermionic ADAPT-VQE requires approximately 2208 CNOT gates for H$_4$ and 28632 for H$_6$, while optimized variants reduce these numbers significantly to 812 CNOT gates for H$_6$~\cite{grimsleyAdaptiveVariationalAlgorithm2019,tangQubitADAPTVQEAdaptiveAlgorithm2021,ramoaReducingResourcesRequired2024}. In contrast, our one or two layers HEA cicruit corresponds to only 108-208 CNOT gates for H$_4$ and 168-324 CNOT gates for H$_6$, depending on the number of HEA layers, respectively,  and using the Hadamard test to measure of diagonal elements of $\boldsymbol{\mathcal{H}}$ and $\boldsymbol{\mathcal{S}}$.

\section{Results}

In this section, we evaluate the performance of IVQE and VQSM algorithms using different proposed cost functions, with the H\(_n\) linear chain as a test case. Note that for H\(_2\), chemical accuracy is achieved after only one iteration, regardless of the cost function used or the bond length.  We employ the STO-3G basis set and the Jordan--Wigner transformation, which requires \(2n\) qubits for H\(_n\), along with an additional ancilla qubit when using the Hadamard test.

\subsection{Convergence with respect to the HEA circuit depth}

\begin{figure*}
  \centering
  {\sffamily {\scriptsize (a)}}\includegraphics[width=0.99\columnwidth]{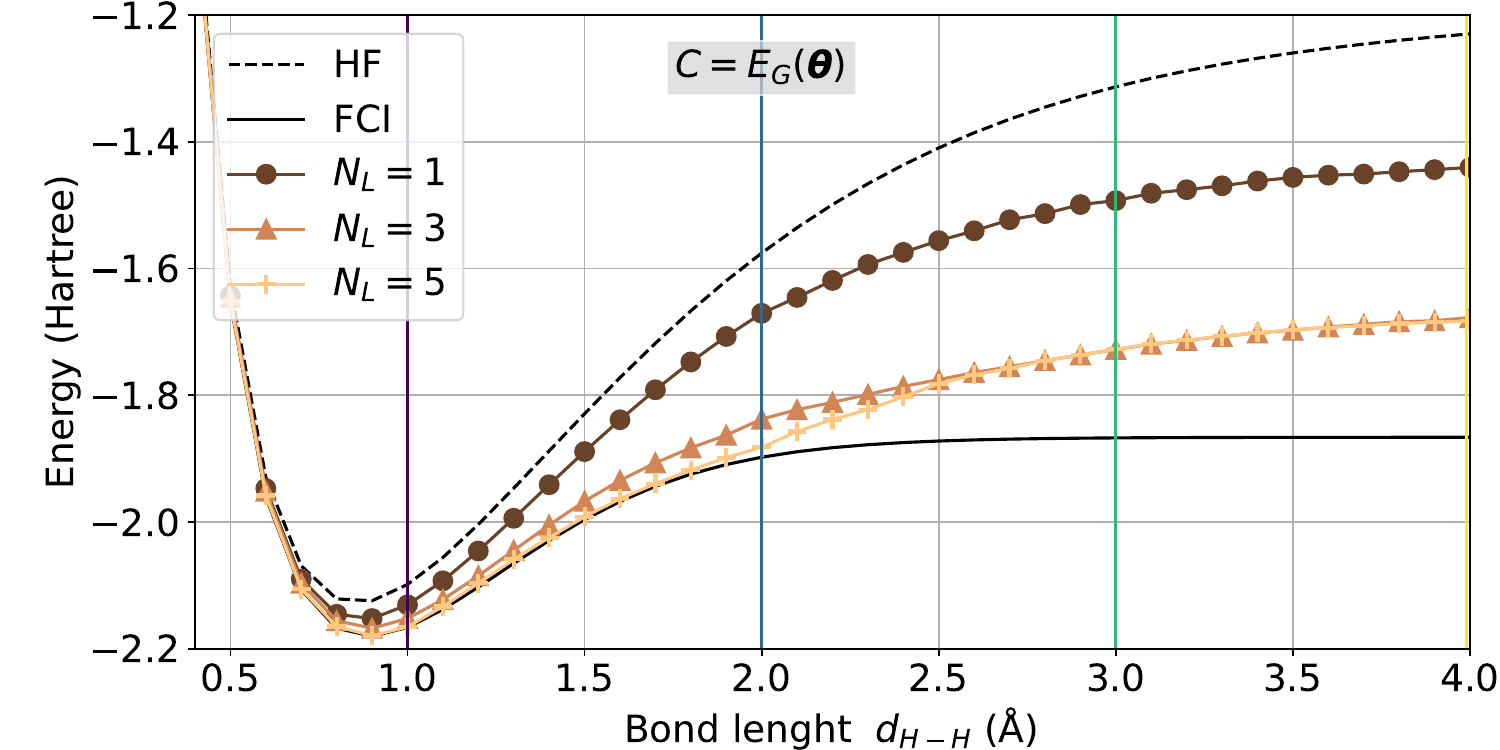}
  {\sffamily {\scriptsize (c)}}\includegraphics[width=0.99\columnwidth]{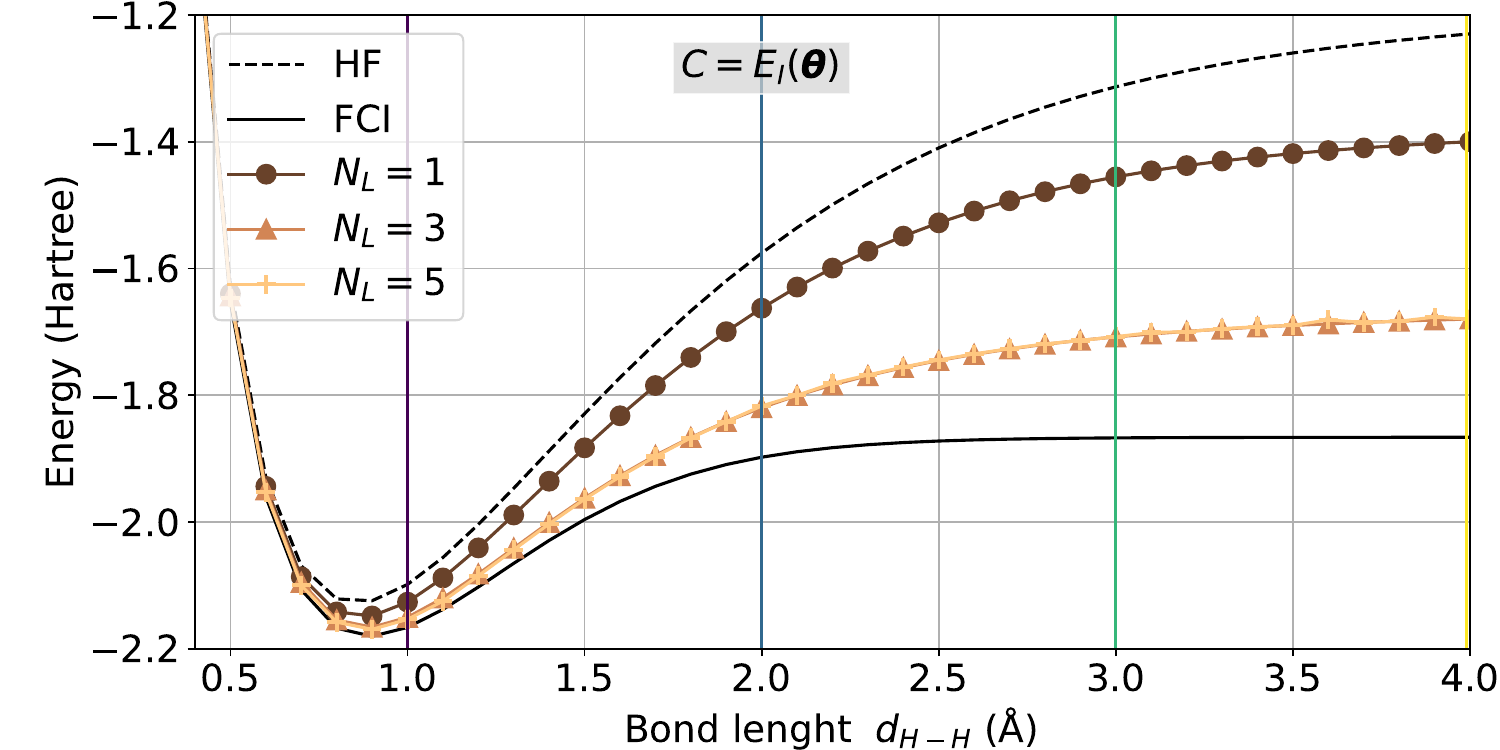} \\
  {\sffamily {\scriptsize (b)}}\includegraphics[width=0.99\columnwidth]{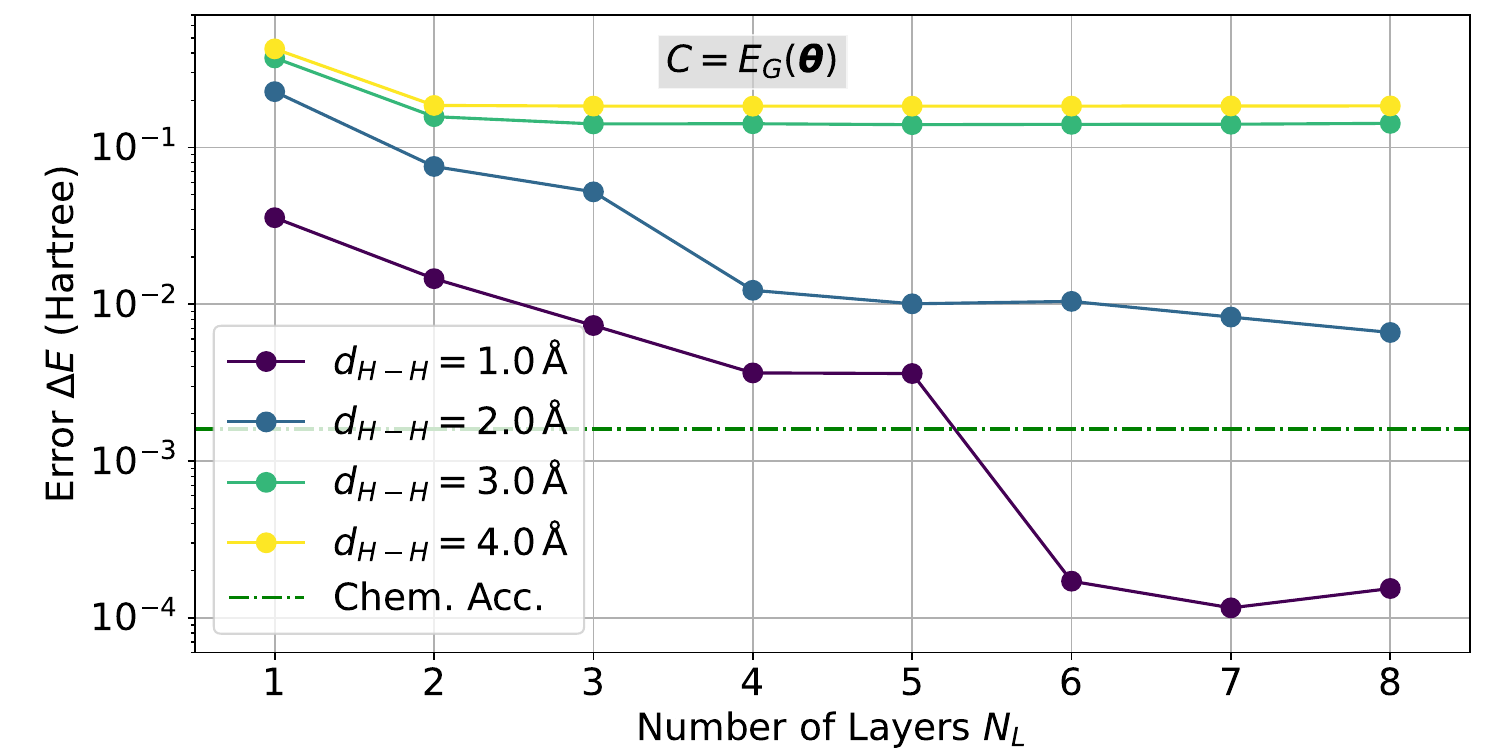}
  {\sffamily {\scriptsize (d)}}\includegraphics[width=0.99\columnwidth]{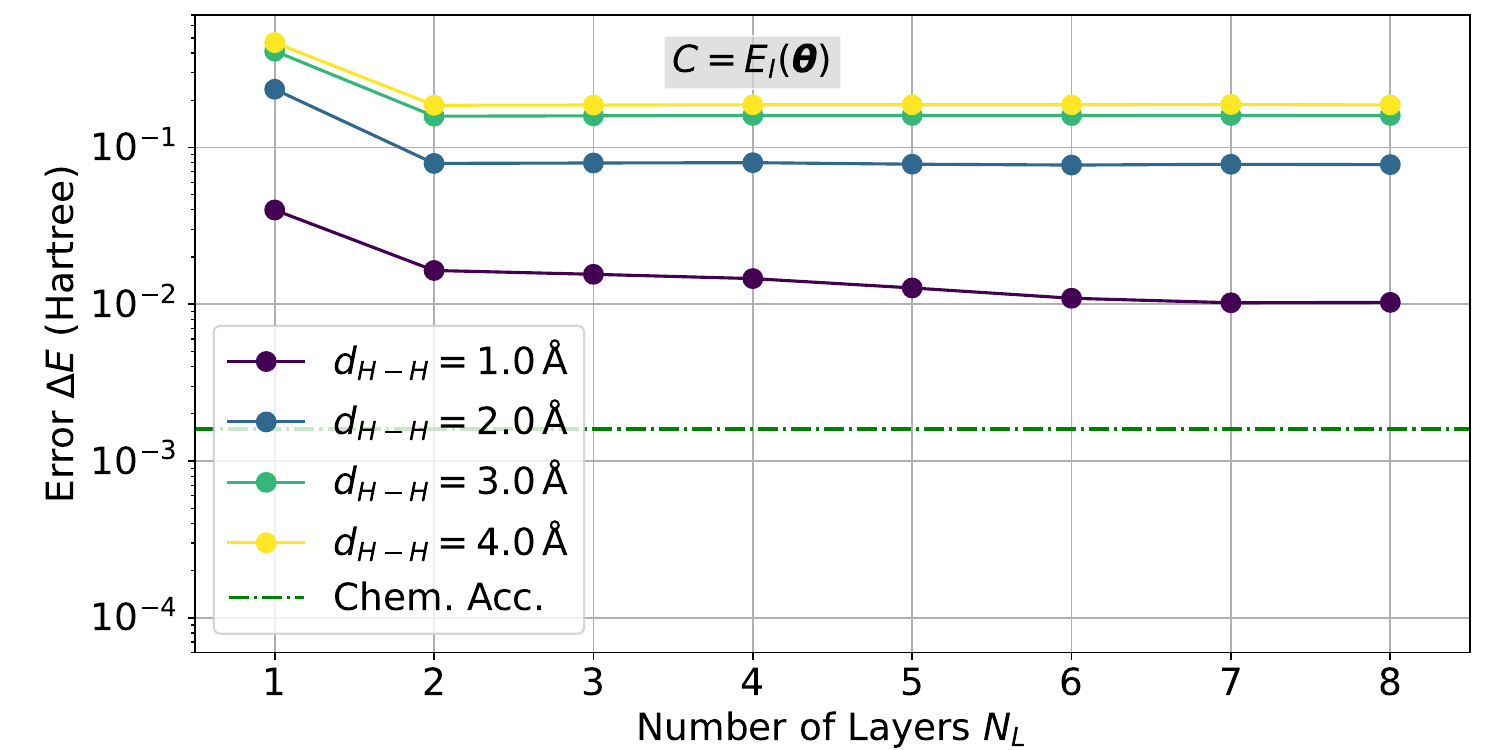}
  \caption{{\it Ground-state energy at the first iteration.} Panels (a) and (b):
    ground-state energy (in Hartree) as a function of the interatomic distance $d_{\rm H-H}$ (in Angstr\"om) for different circuit depths corresponding to  $N_L =1, 3$, or $5$ layers in the HEA using $E_G({\bm\theta})$ and $E_I({\bm\theta})$ cost functions, respectively.
    Results are compared with values obtained using the Hartree--Fock approximation and FCI method.
    Panels (c) and (d): absolute error $\Delta E = E(\bm\theta^*) - E^{\rm FCI}$ (in Hartree) as a function of the number of layers in the HEA for selected interatomic distances of  $d_{\rm H-H} = 1.0, 2.0, 3.0$, and $4.0 {\mathrm \AA}$ using $E_G({\bm\theta})$ and $E_I({\bm\theta})$ cost functions, respectively.}
  \label{fig:fig_1_E}
\end{figure*}

\begin{figure*}
  \centering
  {\sffamily {\scriptsize (a)}}\includegraphics[width=0.99\columnwidth]{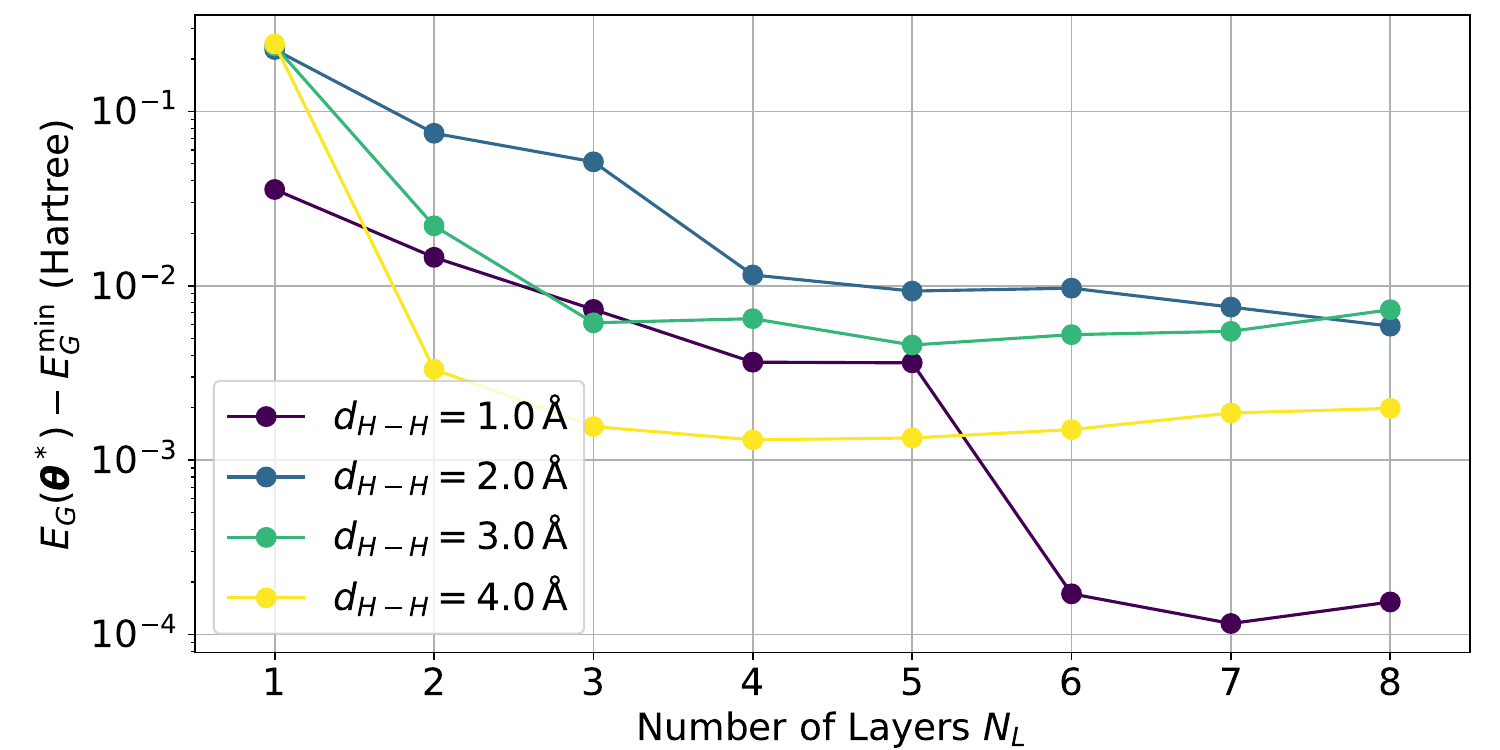}
  {\sffamily {\scriptsize (c)}}\includegraphics[width=0.99\columnwidth]{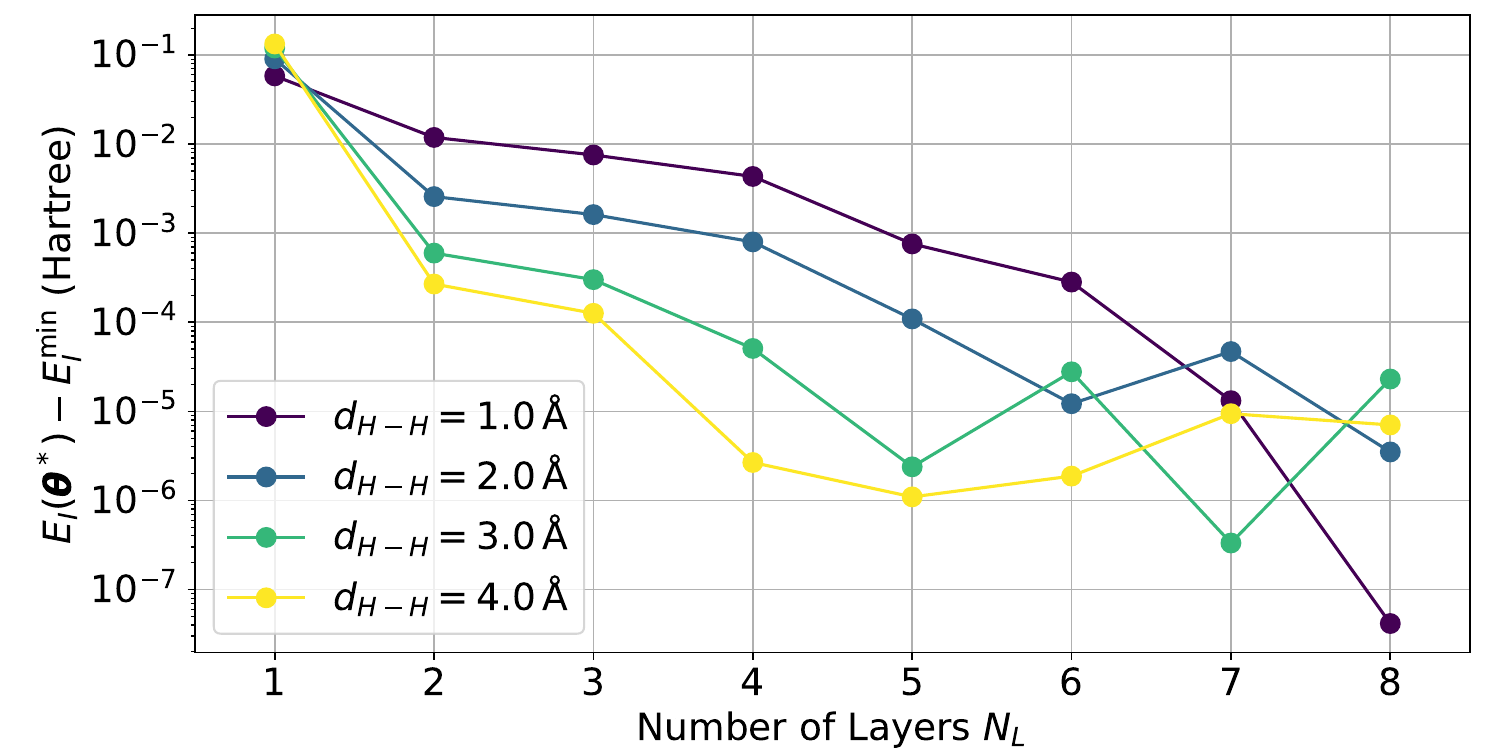}\\
  {\sffamily {\scriptsize  (b)}}\includegraphics[width=0.99\columnwidth]{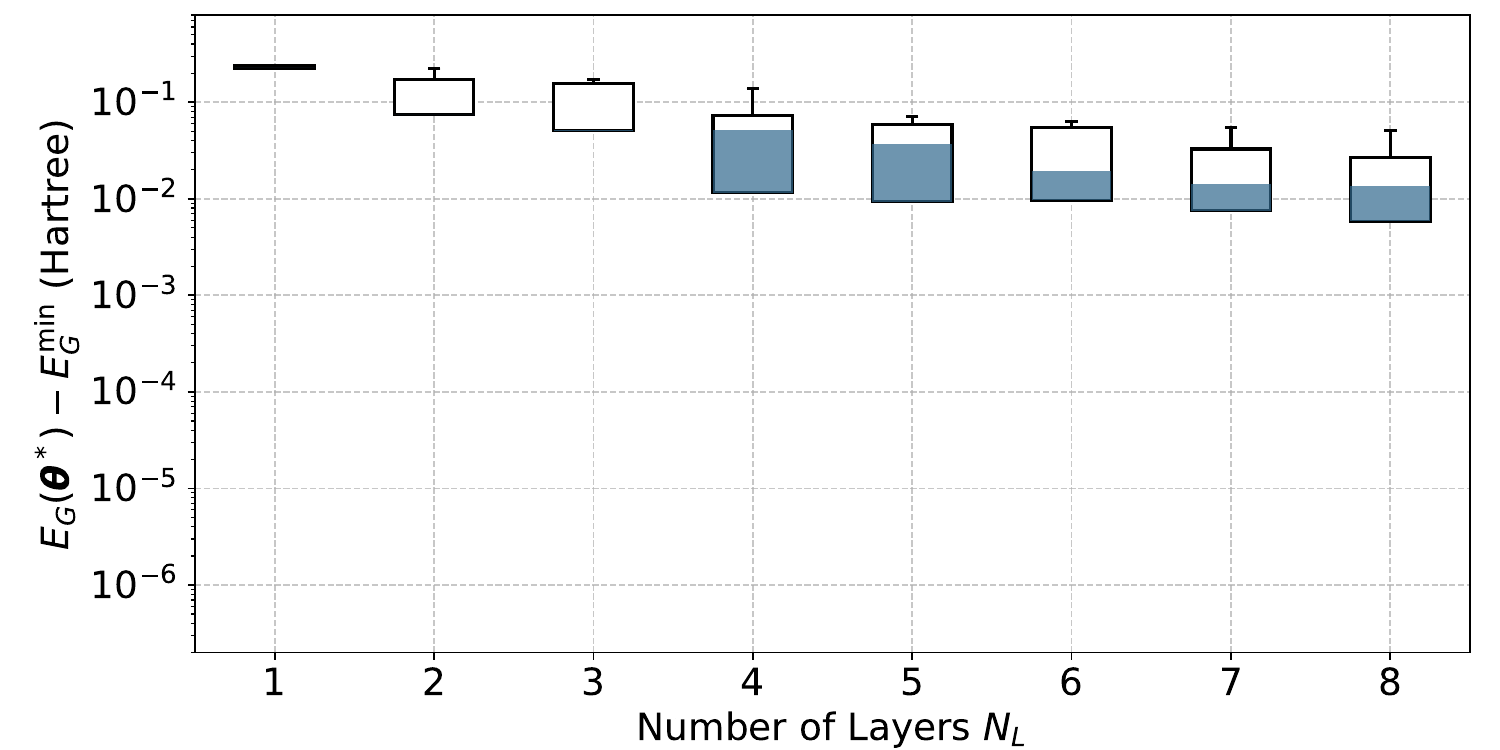}
  {\sffamily {\scriptsize (d)}}\includegraphics[width=0.99\columnwidth]{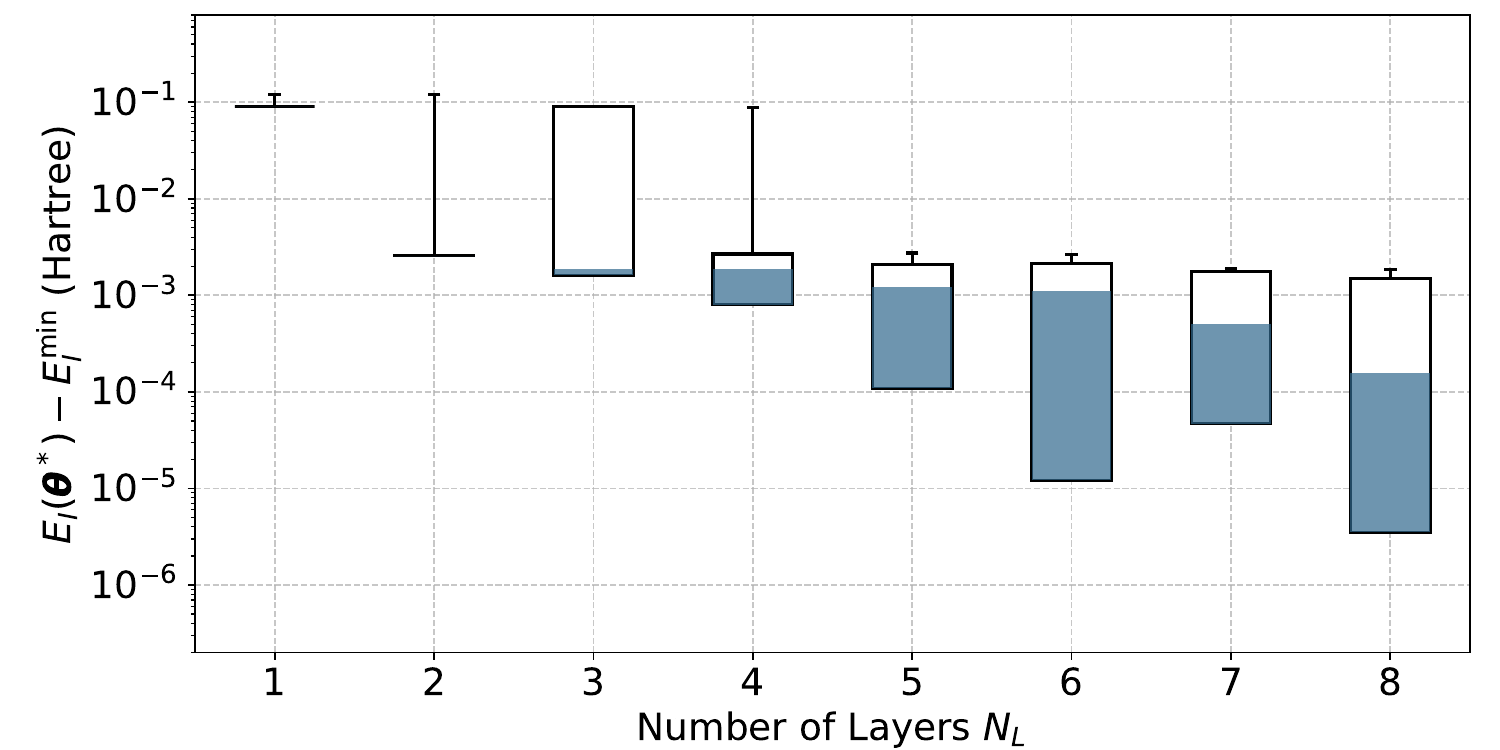}
  \caption{{\it Convergence with respect to the circuit depth and optimization issues at first iteration.}
   Panels (a) and (c): Difference between the cost function $C(\bm \theta^*)$ (in Hartree) with the optimal value  $C^{\rm min}$, as a function of the number of layers in the HEA for selected interatomic distances of  $d_{\rm H-H} = 1.0, 2.0, 3.0$, and $4.0 {\mathrm \AA}$, using $E_G({\bm\theta})$ and $E_I({\bm\theta})$ cost functions, respectively.
  Panels (b) and (d):  Difference between the cost function $C(\bm \theta^*)$ (in Hartree) with the optimal value  $C^{\rm min}$, as a function of the number of layers in the HEA  with box-plot representing 25\% (colored box), 50\% (empty box) and 75\% (whiskers) of optimization outcomes for the cost function when starting with different random  variational parameters. Results are given as a function of the number of layers in the HEA for interatomic distances of  $d_{\rm H-H} = 2.0 {\mathrm \AA}$, using $E_G({\bm\theta})$ (b) and $E_I({\bm\theta})$ (d) cost functions.} 
  \label{fig:fig_2_depth}
\end{figure*}
We first propose to focus on the results obtained at the first iteration, where there are no differences between the IVQE and VQSM approaches. Indeed, at this stage, the constructed subspace of VQSM matches the two-level system of IVQE. This allows us to carefully investigate the influence of the HEA circuit depth and to demonstrate that short circuits can be used. In parallel, we study the influence of choosing  $E_G$ or $E_I$ in particular by showing that the topology of the energy landscape -- and thus the ease of performing classical optimization -- depends drastically on the choice of cost function.

In Fig.~\ref{fig:fig_1_E} we present the ground-state energy of H$_4$ chains as a function of the interatomic distance $d_{\rm H-H}$ together with the error $\Delta E = E(\bmtheta^*) - E^{\rm FCI}$ with respect with the corresponding full configuration interaction (FCI) energy
as a function of the number of layers  $N_L$. 
Results for $E_G(\bm\theta)$ (panels (a) and (c)) are compared  with those of $E_I(\bm\theta)$ (panels (b) and (d)). 
The optimizations are performed using the COBYLA method, either until convergence is reached or until the process is arbitrarily stopped after 150{,}000 cost function evaluations, a setup consistently used for all results presented throughout the manuscript.
It shows that using the $C(\bm\theta) =E_G(\bm\theta)$, for small interatomic distances (i.e., $d_{\rm H-H} \leq 1.9~{\rm \AA}$), the energy closely matches the FCI energy when $N_L = 5$ and that  $\Delta E$ systematically decreases when increasing the circuit depth. 
More precisely, in that case the chemical accuracy can be achieved using only a few layers (e.g., six layers for $d_{\rm H-H} = 1$, as seen in (b). However, when increasing $d_{\rm H-H}$ beyond $1.9~{\rm \AA}$, the ground-state energy appears to converges faster but saturates at a value significantly higher than the FCI energy. As discussed in Sec.~\ref{subsec:Cost}, this behavior results 
from the violation of the \textit{good-guess} assumption -- that is, the HF weight $\omega^2$ in the ground state is lower than 1/2. As shown in Appendix~\ref{app:HFweight}, $\omega^2(\bmtheta^*)$ quickly saturates at a value of 1/2 after only a few layers for $d_{\rm H-H} \geq 2${\AA}. This drawback is overcome by 
starting the iterative scheme, as will be shown later on. 
When using $C(\bm\theta) = E_I(\bm\theta)$  an accurate description of the ground state is not expected in the first iteration since it optimizes the interaction energy with the Hartree--Fock state. It follows that in this case significant deviation with the FCI energy is expected. 
In particular, $\Delta E$ is always higher than 10 mHartree, regardless of the interatomic distance (see panel (d)). Interestingly, the errors on the energy converge rapidly and saturate within only a few HEA layers, suggesting that rather shallow circuits
can be used. Note that as expected, using $E_G(\bmtheta)$ or $E_I(\bmtheta)$ leads to the same saddle point for large \( d_{\rm H-H} \), as suggested by Eq.~(\ref{eq:cost_sim}) since at large $d_{H-H}$, $E_G\sim \sqrt{\widetilde{H}_{01}\widetilde{H}_{10}/\Delta} \propto E_I$. 

Regarding the convergence of the cost function, we show in Fig.~\ref{fig:fig_2_depth} the difference between the cost function $C(\bm\theta^*)$  ($C = E_G$ or $E_I$) obtained withing $N_L$ layers of HEA and its optimal value $C^{\rm min}$ obtained classically by ``brut-force'' minimization of $C({\bf R})$ where ${\bf R} = \mathbb{1} - 2|{\bf v}\rangle \langle {\bf v}|$ is a variational reflection with $|{\bf v}\rangle$ a normalized vector containing the variational parameters, of length 256 for H$_4$.
The minimization of $C({\bf R})$ is numerically efficient and converges rapidly for both cost functions, even though $E_G({\bf R})$ is not linear in ${\bf R}$, unlike $E_I({\bf R})$.
However, When ${\bf R}$ is replaced by a parametrized HEA quantum circuit, the optimization is more tedious. 
To analyze more deeply the optimization properties and the topology of the energy landscape, we conduct a stochastic analysis of the optimization 
process by running multiple optimizations with different random initial parameters. The results are shown in Fig.~\ref{fig:fig_2_depth}(b) and (d), where a box plot illustrates the difference 
between the optimized cost function $C(\bmtheta^*)$ and its exact optimal value $C^{\rm min}$ for $d_{\rm H-H} = 2~{\rm \AA}$ with respect to different number of layers $N_L$. The whiskers, empty boxes, and filled boxes highlight the regions containing the 75\%, 50\%, and 10\% lowest optimized 
cost function values $C(\bmtheta^*)$, respectively. Overall, the results indicate that the energy landscape is far from convex, making it challenging to reach the global minimum -- particularly as the circuit depth increases. 
This suggests that global optimization approaches such as Basin-hopping~\cite{burtonExactElectronicStates2023} are required to reach the global minimum, rather than local optimizers such as gradient-based.
Notably, the convergence rate appears to be significantly accelerated when using $E_I(\bmtheta)$ instead of $E_G(\bmtheta)$.
Consequently, the mapping of \( {\bf R} \) into a unitary parametrized quantum circuit, especially using HEA-type circuits, is itself non-convex and exhibits parameter redundancy. This makes the optimization of \( E_I(\bmtheta) \) still a complex task, particularly as the number of layers and variational parameters increase.
Since \(E_I(\bm{\theta})\) is expected to be numerically more stable and less prone to optimization issues than \(E_G(\bm{\theta})\), we focus on it for the remainder of the manuscript. In the following, we address both the optimization challenges and the noise inherent in deep HEA circuits by trading circuit depth for a higher number of iterations. We show that increasing the number of iterations can yield better results even with shallow circuits composed of only one or two layers.

  In summary, while optimization challenges are inherent to variational algorithms, the use of shallow circuits and a symmetry-preserving cost function is believed to enhances convergence and stability. Nonetheless, we acknowledge the limitations of hardware-efficient ansätze and anticipate that incorporating more expressive, problem-tailored circuits or adaptive strategies could further improve optimization performance in future work. Moreover, the iterative framework is expected to mitigate their impact by transforming a difficult global optimization into a sequence of smaller, more manageable subproblems.

\subsection{Trading number of layers for number of iterations}

\begin{figure*}
  \centering
  {\sffamily {\scriptsize (a)}}\includegraphics[width=0.99\columnwidth]{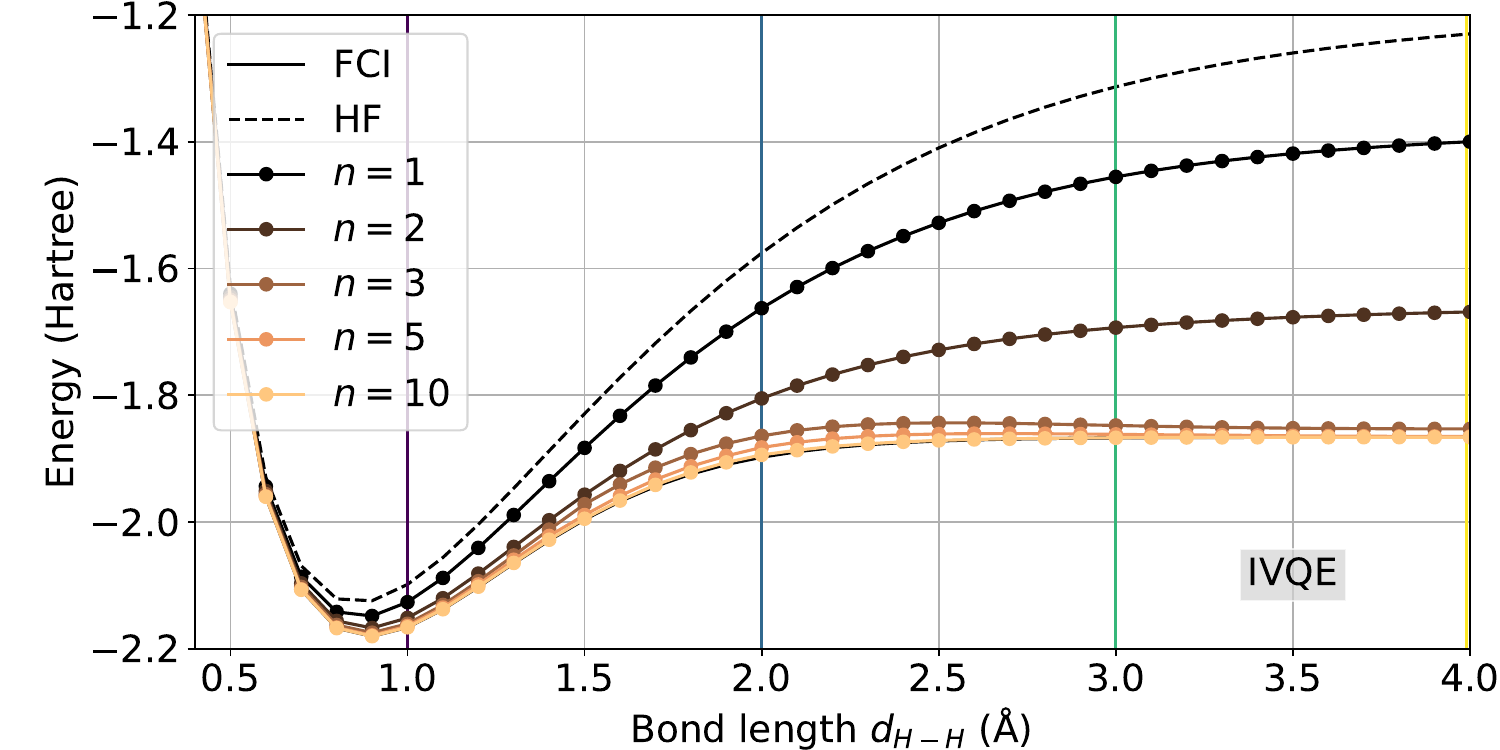}
  {\sffamily {\scriptsize (c)}}\includegraphics[width=0.99\columnwidth]{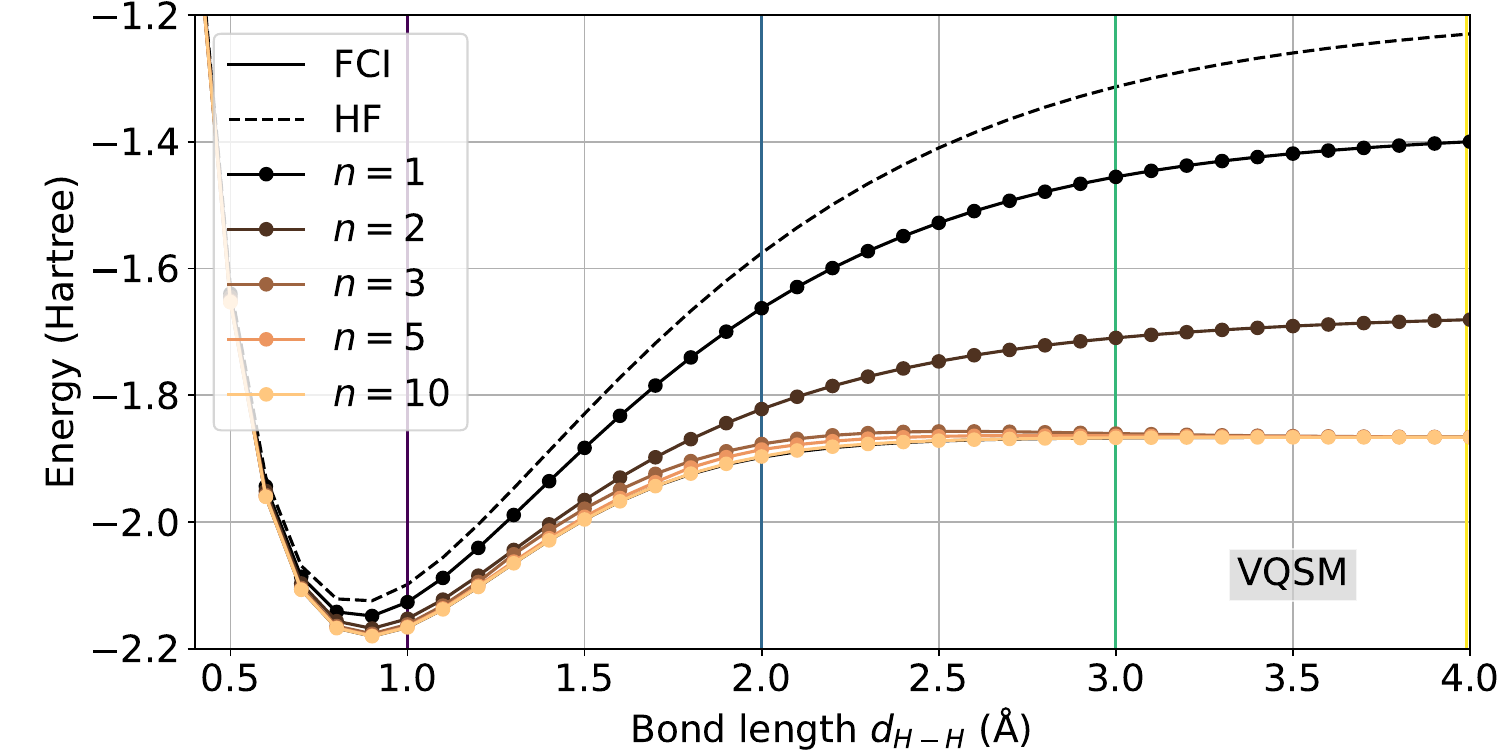} \\
  {\sffamily {\scriptsize (b)}}\includegraphics[width=0.99\columnwidth]{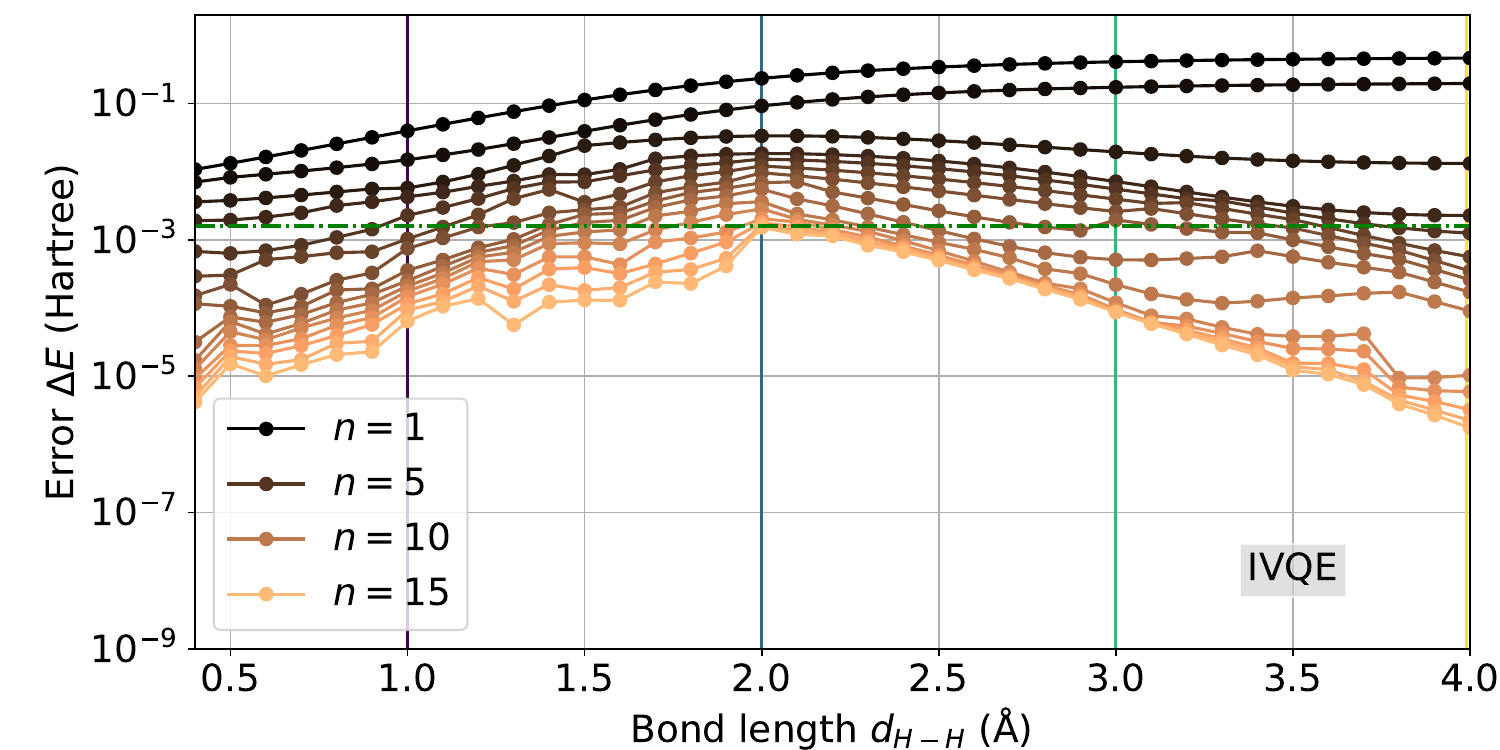}
  {\sffamily {\scriptsize (d)}}\includegraphics[width=0.99\columnwidth]{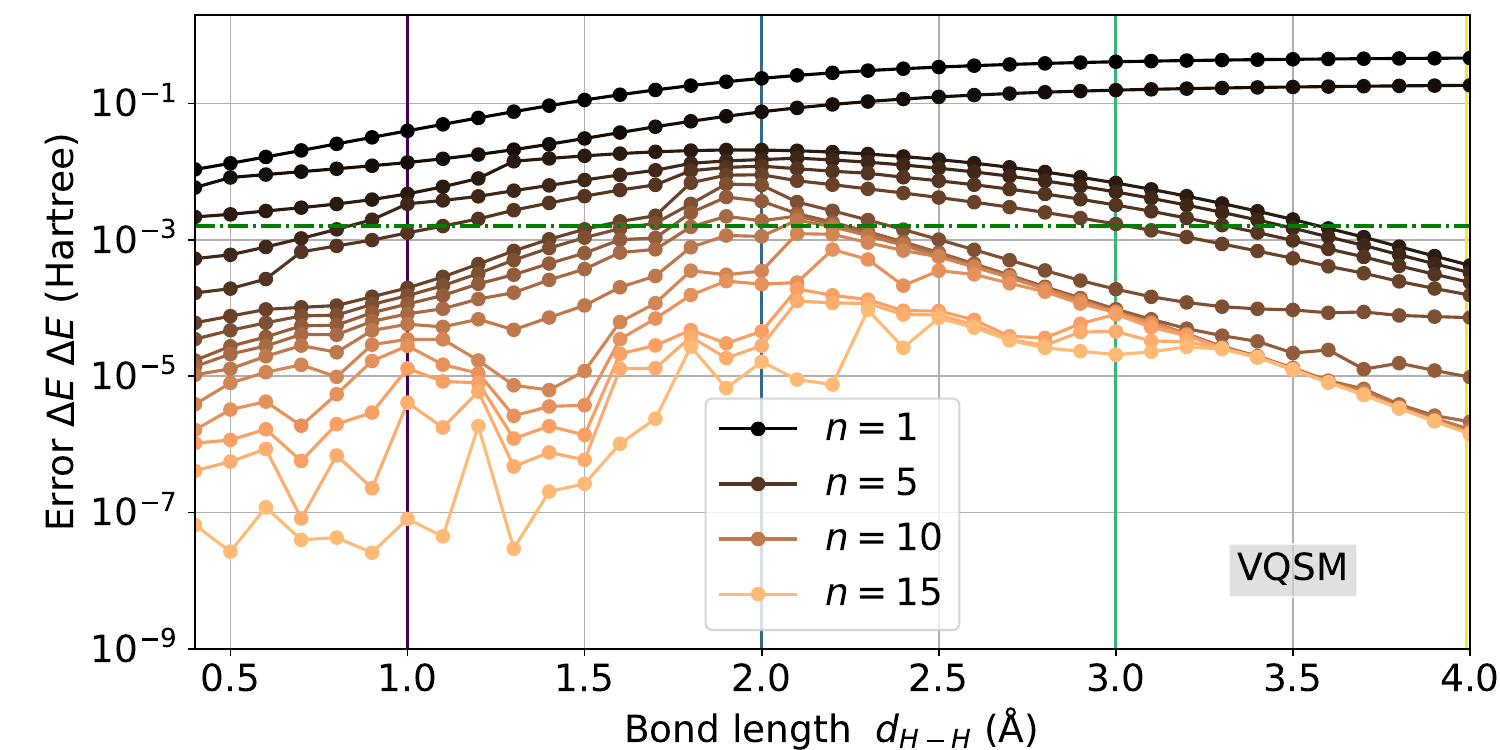}
  \caption{{\it Single layer HEA for linear  $H_4$ molecule IVQE vs VQSM.}
 Ground-state energy (a) and corresponding error $\Delta E$ (b), in Hartree, as a function of the interatomic distance \( d_{\mathrm{H\text{-}H}} \) (in Ångström), using the cost function \( E_I(\bm{\theta}) \) and within the IVQE algorithm. In panel (a), results are shown at iterations \( n = 1, 2, 3, 5, \) and \( 10 \). In panel (b), results are shown for \( n = 1\text{--}15 \), with color shading ranging from dark (early iterations) to light (later iterations), as illustrated for selected iterations in the legend. Results are compared to the Hartree--Fock approximation and exact diagonalization (FCI). The horizontal green dot-dashed line in panel (b) indicates the chemical accuracy threshold (1.6 mHartree). Panels (c) and (d) show the same quantities as (a) and (b), respectively, but for the VQSM algorithm.}
  \label{fig:ResultsIVQE_VQSM}
\end{figure*}

Fig.~\ref{fig:ResultsIVQE_VQSM} presents the results of the IVQE and VQSM algorithms for computing the ground-state properties of the H$_4$ linear chain as a function of the interatomic distance $d_{\rm H-H}$. 
The study uses a single-layer HEA and compares the results to the HF approximation and FCI solutions. 
The top panels (a) and (c) shows the ground-state energy (in Hartree) as a function of interatomic distance for different iteration numbers (1, 2, 3, 5, 10), using the $E_I(\bmtheta)$ cost function. 
Initially, for small iteration counts, both IVQE and VQSM methods deviate significantly from the FCI reference, especially near the dissociation regime. However, as the number of iterations increases, the results get closer to the exact energy. From a number of iterations $n \geq 10$, 
the curve aligns well with the FCI solution, demonstrating that both IVQE and VQSM refine their estimates effectively.
Panels (b) and (d) quantify the error on the energy $\Delta E$ of IVQE and VQSM, respectively, plotting the deviation from the FCI results. $\Delta E$  consistently decreases as the number of iterations $n$ increases, indicating that both methods converge monotonically with respect to the number of iterations. The improvement in accuracy is more significant at small and large bond lengths, with errors falling below chemical accuracy (1.6 mHartree) after a few iterations. 
This suggests that both methods efficiently capture electronic structure and correlation effects, even at large distances.  However, residual errors larger than the chemical accuracy is persisting at intermediate regime $d_{\rm H-H} \sim 2.0-3.0$~\AA~using IVQE, even after 15 iterations.
A key difference between IVQE and VQSM emerges in their convergence behavior and efficiency. Both methods systematically improve with more iterations, but VQSM converges faster, especially for larger bond lengths. For instance, chemical accuracy is reached after $n= 6, 9, 15$, and $5$ iterations for $d_{\rm H-H} = 1.0, 2.0, 3.0$ and $4.0$\AA, respectively, for IVQE ; compared to $n=5, 7, 10$, and $3$ iterations for VQSM. Therefore, the subspace expansion in VQSM provides a more efficient algorithm, accelerating convergence to the exact solution, at the expense of larger subspace Hamiltonian $\boldsymbol{\mathcal{H}}^{(n)}$ to be diagonalized classically. 
Overall VQSM demonstrates superior convergence properties, highlighting its potential as a more efficient variational algorithm for electronic structure calculations. Accordingly, we focus on VQSM associated with $E_I(\bm \theta)$ as a cost function in the rest of the manuscript.
\begin{figure}
  \centering
  {\sffamily {\scriptsize  (a)}}\includegraphics[width=0.99\columnwidth]{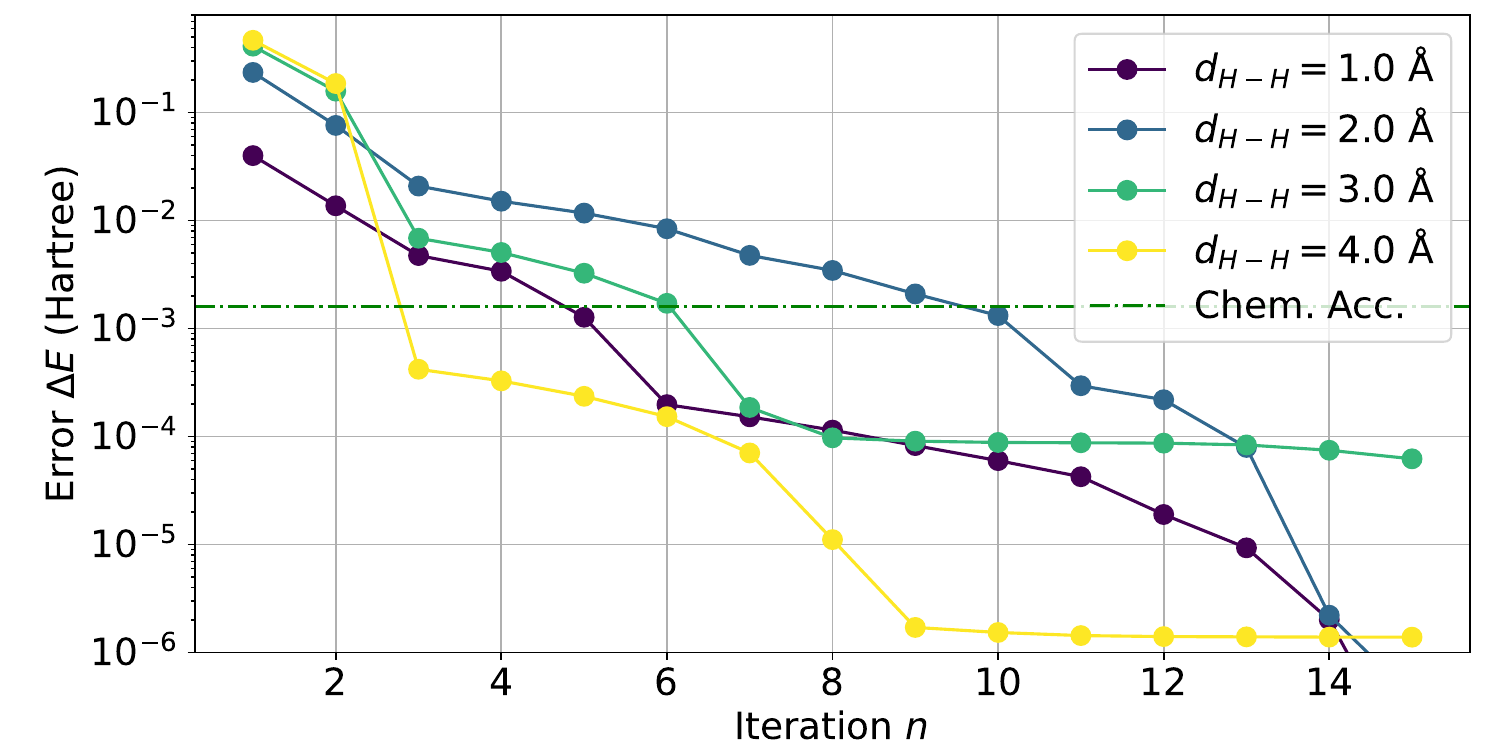}
  {\sffamily {\scriptsize (b)}}\includegraphics[width=0.99\columnwidth]{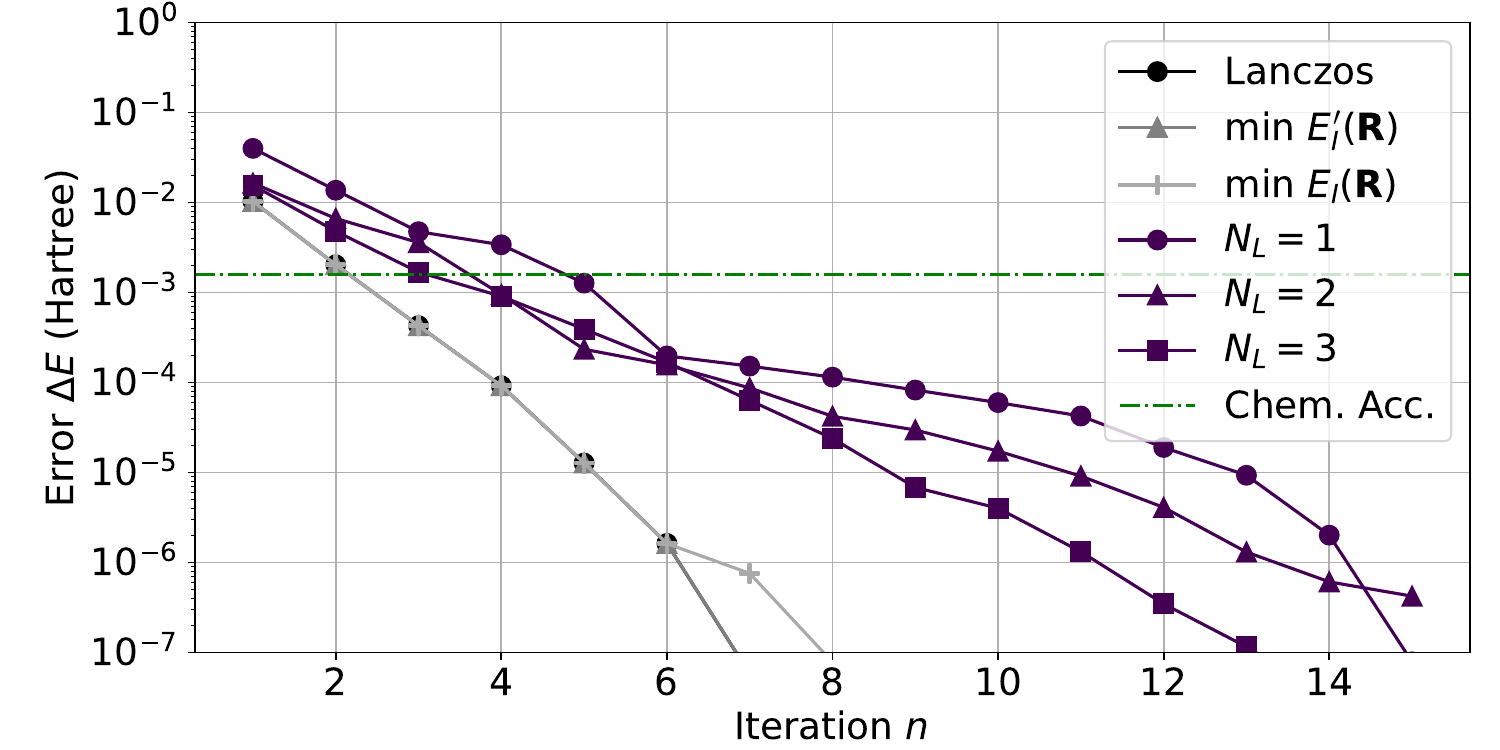}
  {\sffamily {\scriptsize (c)}}\includegraphics[width=0.99\columnwidth]{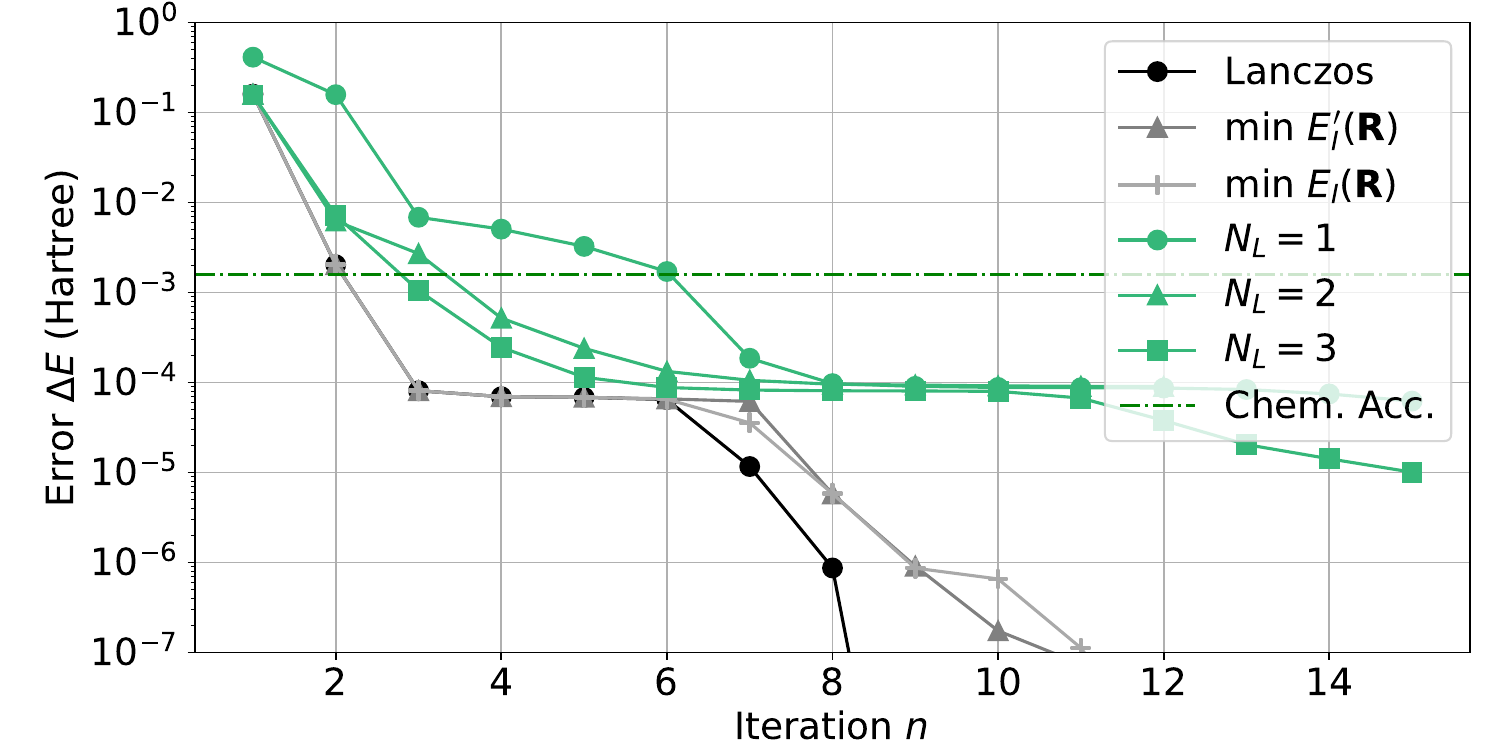}
  \caption{{\it Convergence rate of VQSM for the linear $H_4$ molecule:} 
(a) Energy error $\Delta E$ (in Hartree) as a function of the number of layers $n$, for various bond lengths $d_{\rm H\text{-}H} = 1.0$, $2.0$, $3.0$, and $4.0$~\AA{} (see legend), using the cost function $E_I(\bm{\theta})$ and a single-layer HEA. 
(b) and (c): same as (a) for $d_{\rm H\text{-}H} = 1.0$ (b) and $d_{\rm H\text{-}H} = 3.0$ (c). In (b) and (c), results are shown for HEAs with 1, 2, and 3 layers (colored curves with symbols, see legend) and are compared with classical methods shown using shade of gray. In particular, results obtained using the standard Lanczos algorithm are shown in black, while results from classical minimization of $E_I'(\mathbf{R})$ and $E_I(\mathbf{R})$ are shown in dark gray and light gray, respectively.}
  \label{fig:Results_EI}
\end{figure}
\subsection{Convergence rate of VQSM versus the standard Lanczos algorithm}

\begin{figure*}
  \centering
  {\sffamily  {\scriptsize (a)}}\includegraphics[width=0.99\columnwidth]{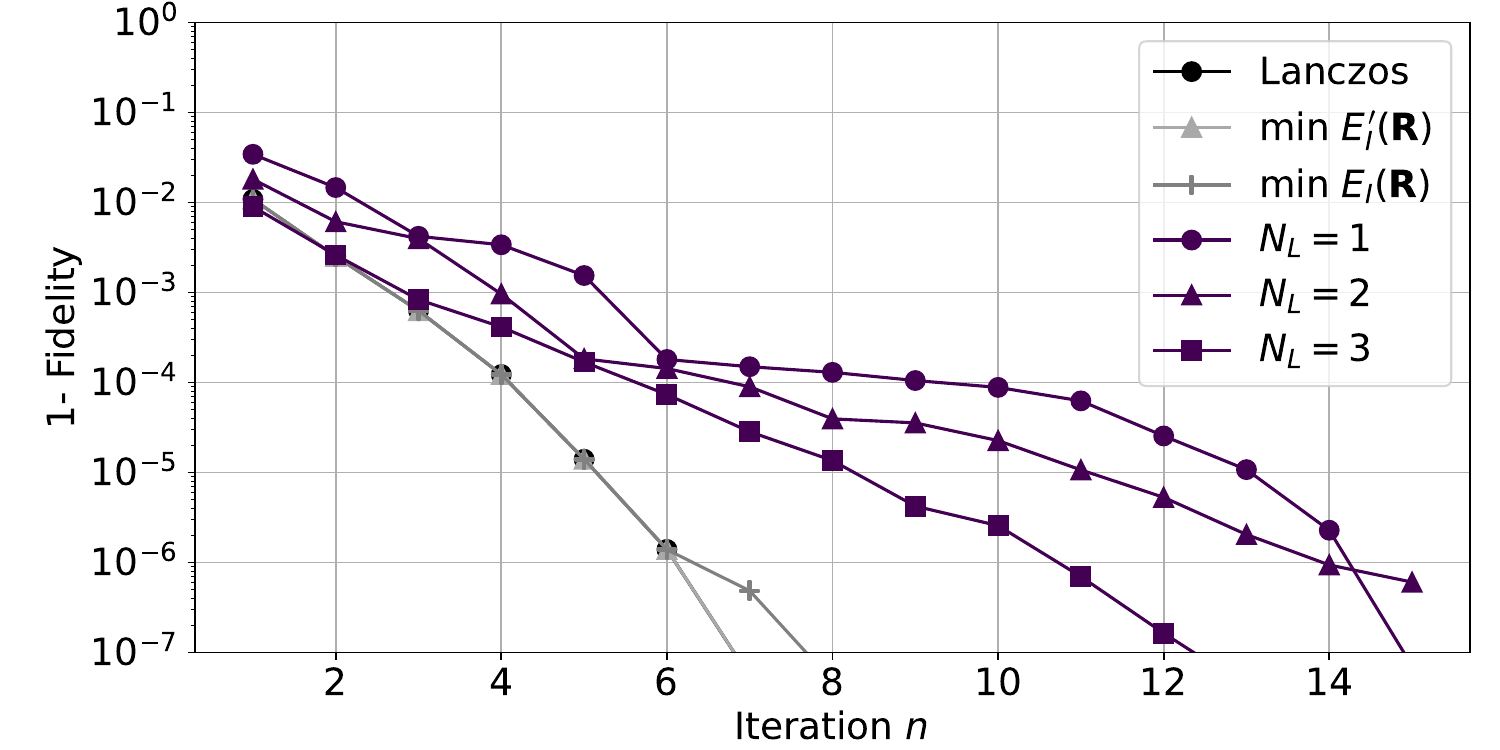}
  {\sffamily  {\scriptsize (c)}}\includegraphics[width=0.99\columnwidth]{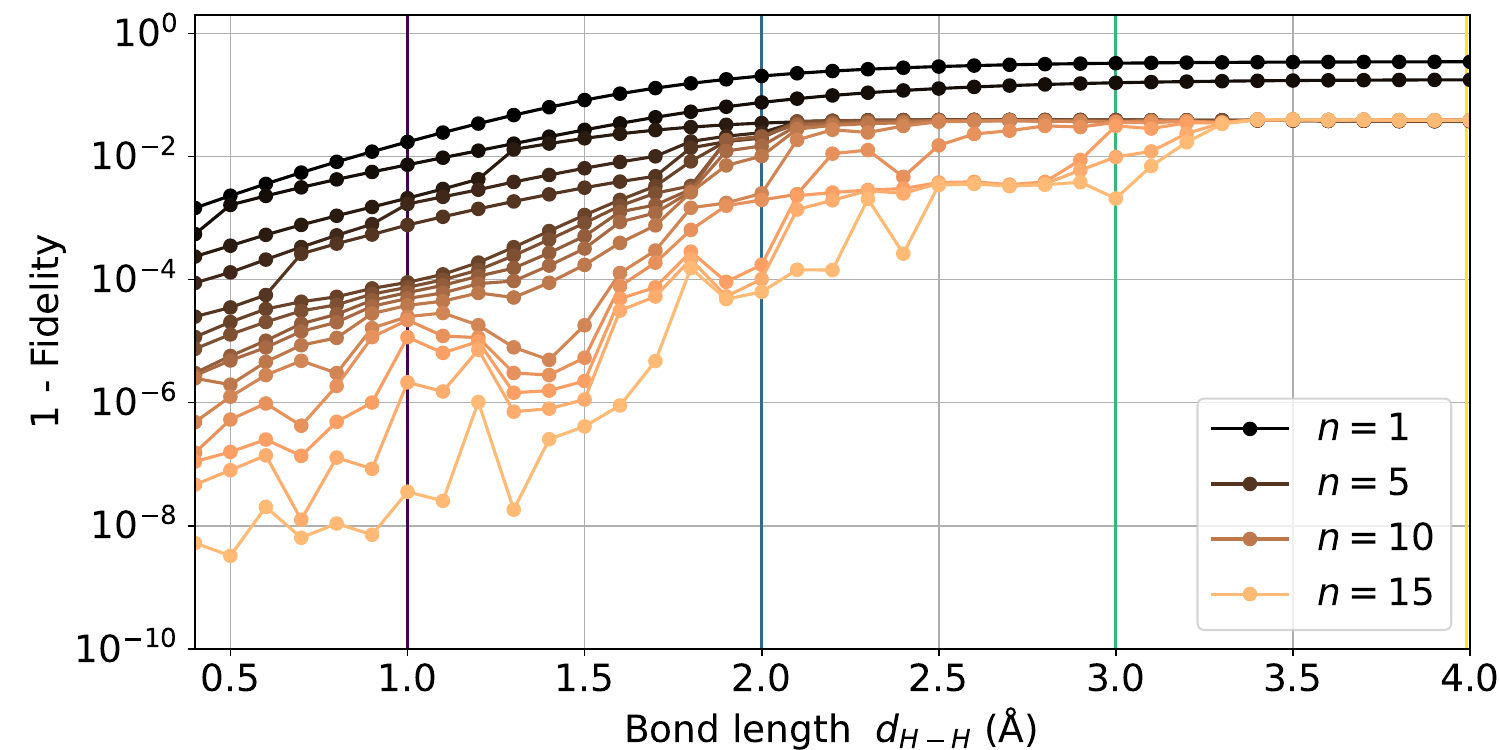}\\
  {\sffamily  {\scriptsize (b)}}\includegraphics[width=0.99\columnwidth]{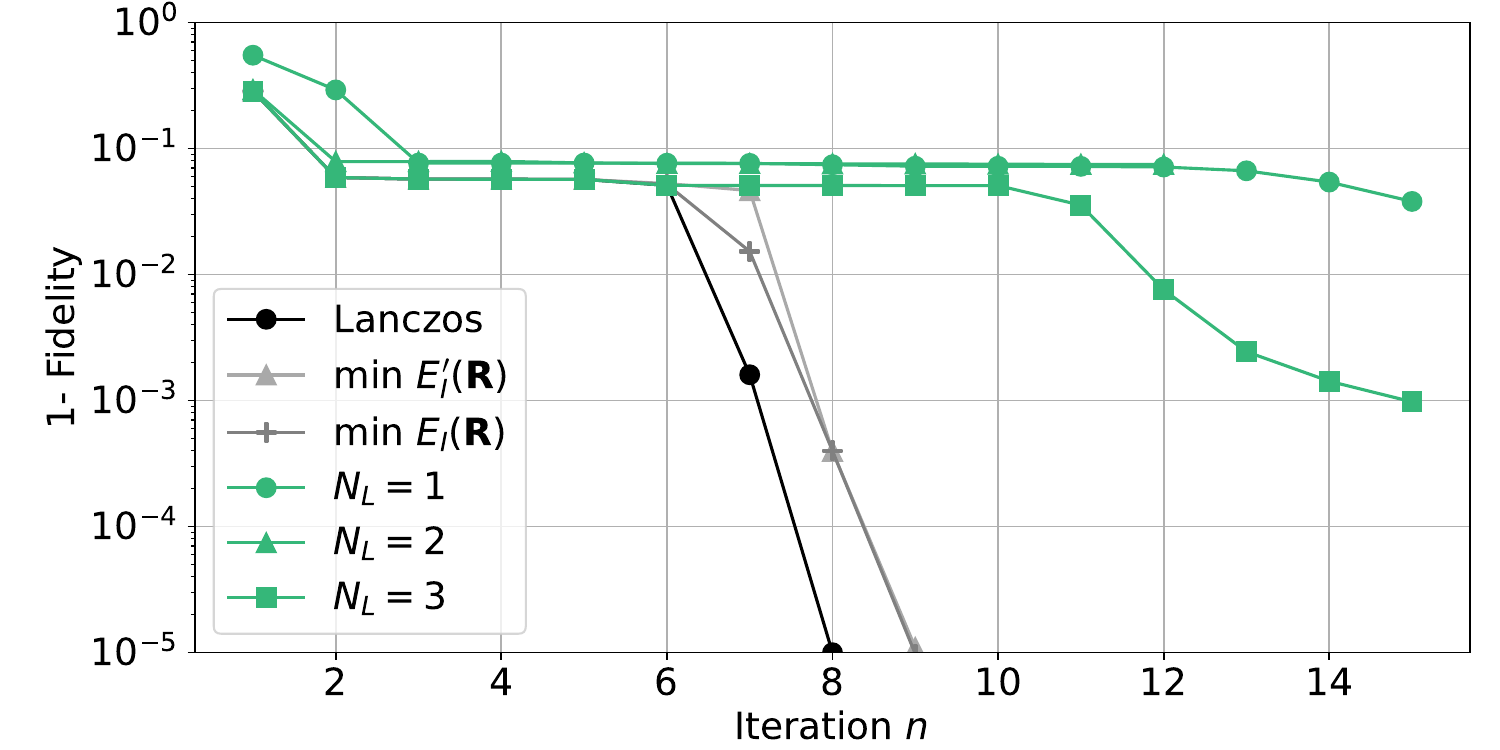}
  {\sffamily  {\scriptsize (d)}}\includegraphics[width=0.99\columnwidth]{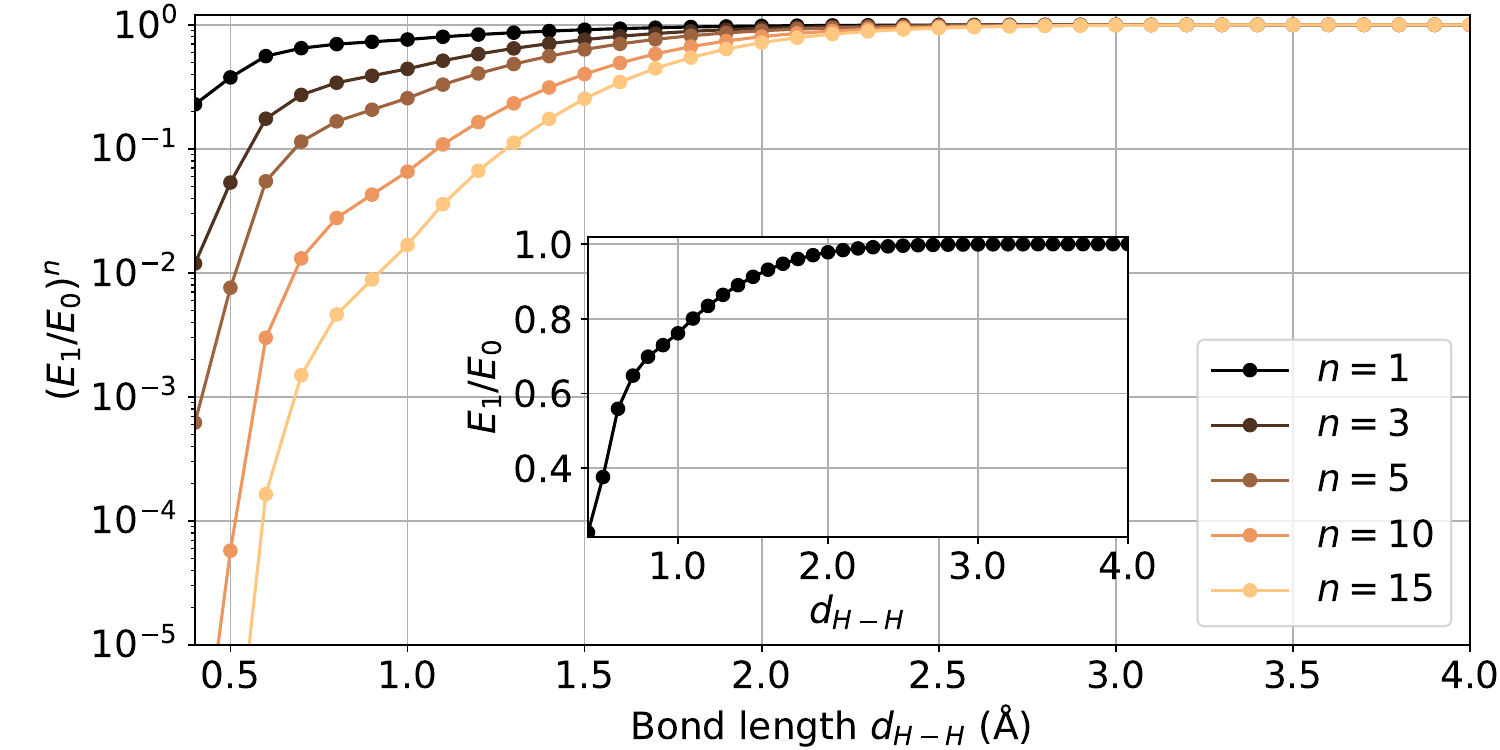}
  \caption{\textit{Convergence rate of the Fidelity using VQSM for the linear H$_4$ chain with $E_I(\bm{\theta})$ as the cost function.}
Panels (a) and (b): Fidelity as a function of the iteration number $n$ at bond lengths $d_{\rm H-H} = 1.0$\AA\ and $3.0$\AA, respectively. 
The colored lines refer to VQSM results obtained for different HEA circuit depths having $N_L = 1, 2$ and $3$ layers, distinguished by the different symbols as specified in the  legend. 
They  are compared with values obtained using Lanczos method (black curve) or by
brute-fore minimization of $E_I'({\bf R})$ (dark gray) and $E_I({\bf R})$ (light gray) with respect to the reflection matrix ${\bf R}$. 
Panel (c): Fidelity as a function of the bond length $d_{\rm H-H}$ (in \AA) at different iterations ($n = 1-15$) with color shading ranging from dark (early iterations) to light (later iterations), as illustrated for selected iterations in the legend.
Panel (d): Powers of the ratio between the ground-state energy $E_0$
and the first excited energy belonging to the same symmetry subspace $E_1$ as a function of $d_{\rm H-H}$ for several value of the iteration $n$. 
The inset highlights the case for $n = 1$.
 }
  \label{fig:VQSM_fid_cvg}
\end{figure*}

In Fig.~\ref{fig:Results_EI}, we analyze the convergence rate of the VQSM algorithm for the linear H$_4$ molecule and compare it with the standard (classical) Lanczos algorithm. As previously discussed in the manuscript, both VQSM using $E_I'({\bm \theta})$ and the Lanczos algorithm are expected to produce the same tridiagonal matrix within the constructed subspace. Consequently, we also expect a geometric convergence for VQSM characterized by an exponential decay of the error: $\varepsilon^{(n)} = \varepsilon^0 r^n$, where $\varepsilon^{(n)}$ denotes the error at iteration $n$, $\varepsilon^0$ is a constant, and $r$ is the convergence rate.
Panel (a) shows the energy error $\Delta E$ obtained using a single-layer HEA as a function of the number of iterations $n$, for different interatomic distances $d_{\rm H-H}$. For small bond lengths ($d_{\rm H-H} \leq 2$\AA),  $\Delta E$ decreases almost linearly (on a logarithmic scale) with the iteration number  exhibiting the expected geometric convergence. The logarithm of the convergence rate $\ln(r)$ can be directly extracted as the slope of the error vs iteration curve when plotted on a logarithmic scale. For $d_{\rm H-H} \leq 2$\AA, we obtain $r \sim 0.5$ using a linear regression, meaning that at each iteration, the error is divided by a factor of two. However, for $d_{\rm H-H} > 2$~\AA, the error initially decreases rapidly in the first few iterations but then slows significantly, leading to a ``plateau-like'' behavior.

The efficiency or convergence rate of the VQSM algorithm is expected to depend on several factors. On one hand, as other Krylov-based algorithm, it is influenced by physical parameters, such as the first eigengap, i.e. the energy gap between the ground and first excited states within the symmetry-constrained Hilbert space. On the other hand, numerical parameters also play a role, including the circuit depth (which determines the ability to span the Fock space) and optimization success (i.e., the ability to reach the global minimum in a complex energy landscape).
To investigate the effect of circuit depth and optimization issue on VQSM convergence, we present in Fig.~\ref{fig:Results_EI}(b) and (c) the ground-state energy error $\Delta E$ as a function of the iteration number for HEA circuits with different layer counts, compared to classical algorithms. We focus on two representative cases: $d_{\rm H-H} = 1.0$~\AA~(b) and $d_{\rm H-H} = 3.0$~\AA~(c).
In parallel, Fig.~\ref{fig:VQSM_fid_cvg} illustrates the convergence of the ground-state fidelity with respect to the FCI solution. Panels (a) and (b) show fidelity versus iteration number for different HEA depths, again for $d_{\rm H-H} = 1.0$~\AA~and $d_{\rm H-H} = 3.0$~\AA, respectively.
The results in Figs.~\ref{fig:Results_EI}(b), \ref{fig:Results_EI}(c), \ref{fig:VQSM_fid_cvg}(a) and \ref{fig:VQSM_fid_cvg}(b) are compared with the classical Lanczos algorithm (black curves). Dark gray curves are obtained by directly minimizing the cost function $E_I' ({\bf R})$ in Eq.~(\ref{eq:cost_Tri}) with respect to a variational reflection ${\bf R} = \mathbb{1} - 2|{\bf v}\rangle \langle {\bf v}|$, where $|{\bf v}\rangle $ is a normalized variational vector of the size of the many-body basis set.
Similarly, light gray curves correspond to results obtained by minimizing $E_I({\bf R})$.
Interestingly, switching from $E_I'(\bmtheta)$ to $E_I(\bmtheta)$ has only a minor impact, meaning that both cost function can  also lead to the same triangular form of the Hamiltonian in the reduced subspace. As expected, minimizing $E_I'(\bmtheta)$ reproduces quantitatively, up to numerical setting, the Lanczos results. First of all, we remark that the ``plateau-like'' behavior, observed in Fig.~\ref{fig:Results_EI}(a) close to dissociation, is also present when using classical methods such as Lanczos. This suggests that the slower convergence near dissociation has a physical origin rather than resulting from optimization issues.
When turning to quantum implementation, replacing the reflection operator by a parameterized unitary HEA circuit (with $N_L = 1, 2, 3$ layers) to minimize $E_I(\bmtheta)$ reduces the convergence rate, since the resulting ``Krylov vector'' is only approximated. For instance, at $d_{\rm H-H} = 1.0$~\AA, the convergence rate for the energy error drops from $r \sim 0.50$ for a single-layer HEA to $r \sim 0.47$ and $r \sim 0.40$ for two- and three-layer HEAs, respectively. Even though the number of layers is increased, it remains far from the convergence rate of the Lanczos algorithm in this case, which shows $r = 0.07$. Note that the convergence trend for fidelity closely follows that of the energy error.

Finally, Fig.~\ref{fig:VQSM_fid_cvg}(c) 
displays the fidelity as a function of $d_{\rm H-H}$ at various iteration steps $n$, which can be compared to powers of the ratio $(E_1/E_0)^n$ in  Fig.~\ref{fig:VQSM_fid_cvg}(d), where $E_0$ and $E_1$ are the ground- and first excited-state energies within the same symmetry subspace. Indeed, for the power method, $r$ is known to be proportional to the ratio $E_1/E_0$. While the convergence rate of the Lanczos algorithm is more precisely governed by the ratio between the first eigengap $E_1 - E_0$ and the diameter of the remaining spectrum~\cite{golubMatrixComputations1996}, the ratio $E_1/E_0$ already provides a rough estimate of the convergence rate. In both cases, for nearly degenerate systems where $E_1/E_0 \rightarrow 1$ (i.e., $E_1 - E_0 \rightarrow 0$), as encountered near dissociation (see the inset of Fig.~\ref{fig:VQSM_fid_cvg}(d)), the convergence becomes slow and can exhibit plateau-like behavior. In contrast, for systems with a large energy gap ($E_1/E_0 \rightarrow 0$ or $E_1 - E_0 \gg 1$), i.e., for shorter $d_{\rm H-H}$, convergence is expected to be fast. This analysis explains the plateau behavior observed in both VQSM and  Lanczos algorithm in Figs.~\ref{fig:Results_EI}(c) and~\ref{fig:VQSM_fid_cvg}(b), and confirms that it arises from physical limitations rather than from optimization artifacts.

\begin{figure}
  \includegraphics[width=1\columnwidth]{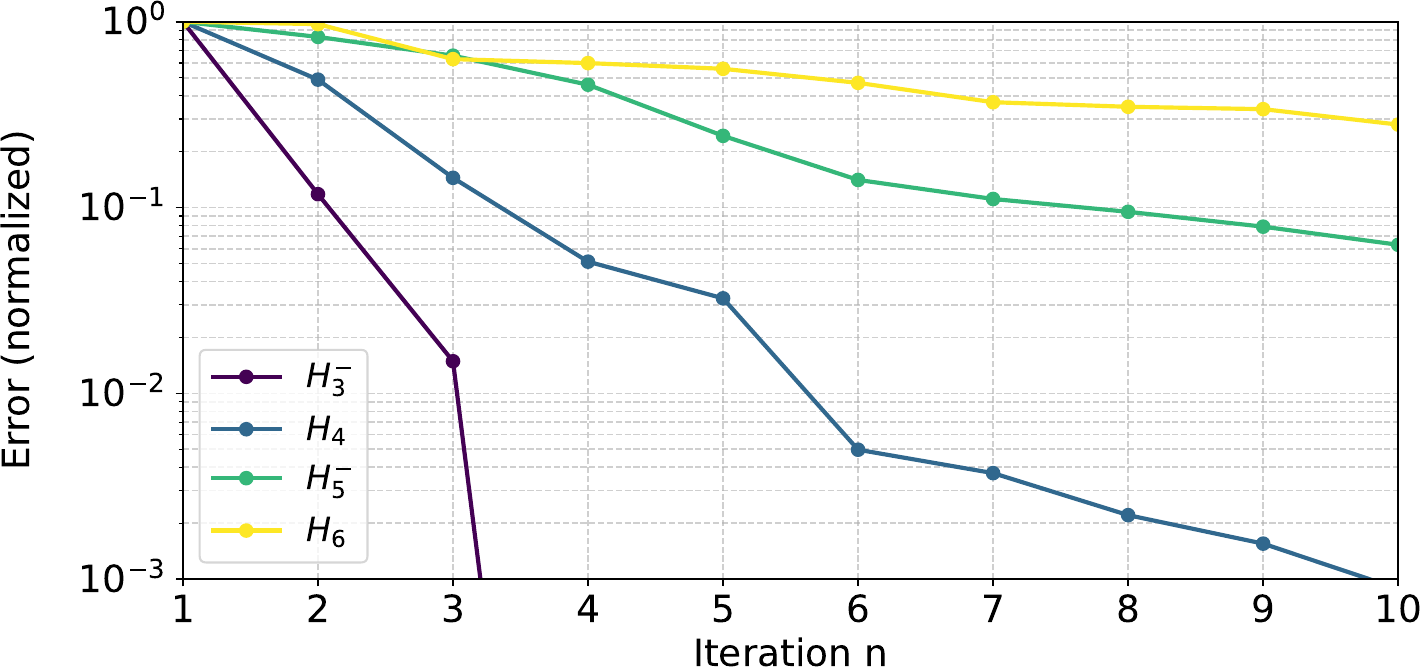}
  \caption{\textit{Geometrical convergence of VQSM on hydrogen chains H$_n$ with $n=3$ to $6$} using a fixed one-layer HEA ansatz and $E_I(\bmtheta)$ Cost Function, for $n=3$ and $5$ we have considered H$_n^-$. For a sake of clarity the relative error is renormalized by the error at the first iteration. 
 }
  \label{fig:VQSM_scaling}
\end{figure}

Altogether, increasing the number of HEA layers $N_L$ improves the convergence rate, although a significantly deep circuit would be required to fully match the performance of the Lanczos algorithm and probably also associated with optimization issues. Nonetheless, a good trade-off can be achieved between (i) the number of iterations, (ii) the circuit depth, and (iii) the occurrence of optimization difficulties to reach a desired level of accuracy. This balance is especially crucial in near-degenerate regimes, where improving circuit expressibility becomes necessary but increasingly costly. In practice, choosing the minimal depth that ensures sufficient convergence while maintaining feasible optimization and coherence constraints on quantum hardware is key for the efficient implementation of VQSM on near-term devices. Finally, as a sake of completeness we show in Fig.~\ref{fig:VQSM_scaling} results for H$_n$ chains $3 \leq n \leq 6$. It shows that as the system size increases, the convergence rate slows down slightly, consistent with a decreasing energy gap and the limited ansatz expressivity. However, the geometric convergence is preserved, demonstrating the robustness of the VQSM as variational Krylov-based approach across different system sizes.

\subsection{Beyond Ground-State Properties}

\begin{figure}
\includegraphics[width=1\columnwidth]{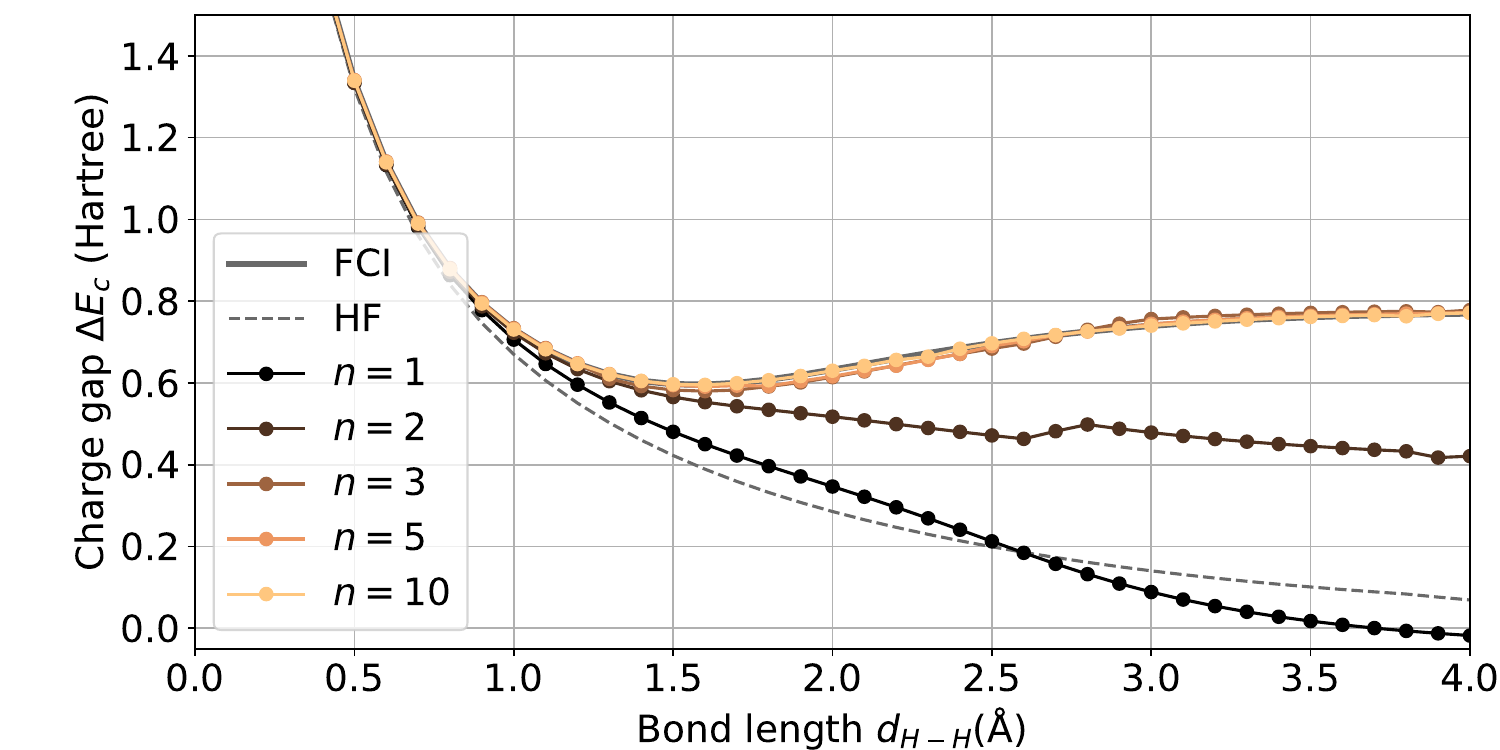}
\caption{\textit{VQSM Band Gap with $E_I(\bmtheta)$ Cost Function and a Single Layer HEA.}  Charge gap $\Delta E_c = E({\rm H}_4^+) + E({\rm H}_4^-) - 2 E({\rm H}_4)$
as a function of the interatomic distance $d_{\rm H-H}$ (in Å) at different iterations ($n=1, 2, 3, 5$, and $10$). Results are compared with values obtained using the Hartree--Fock approximation and exact diagonalization (FCI).
   }
  \label{fig:VQSM_ChargeGap}
\end{figure}
\begin{figure}
  {\sffamily {\scriptsize (a)}}\includegraphics[width=0.99\columnwidth]{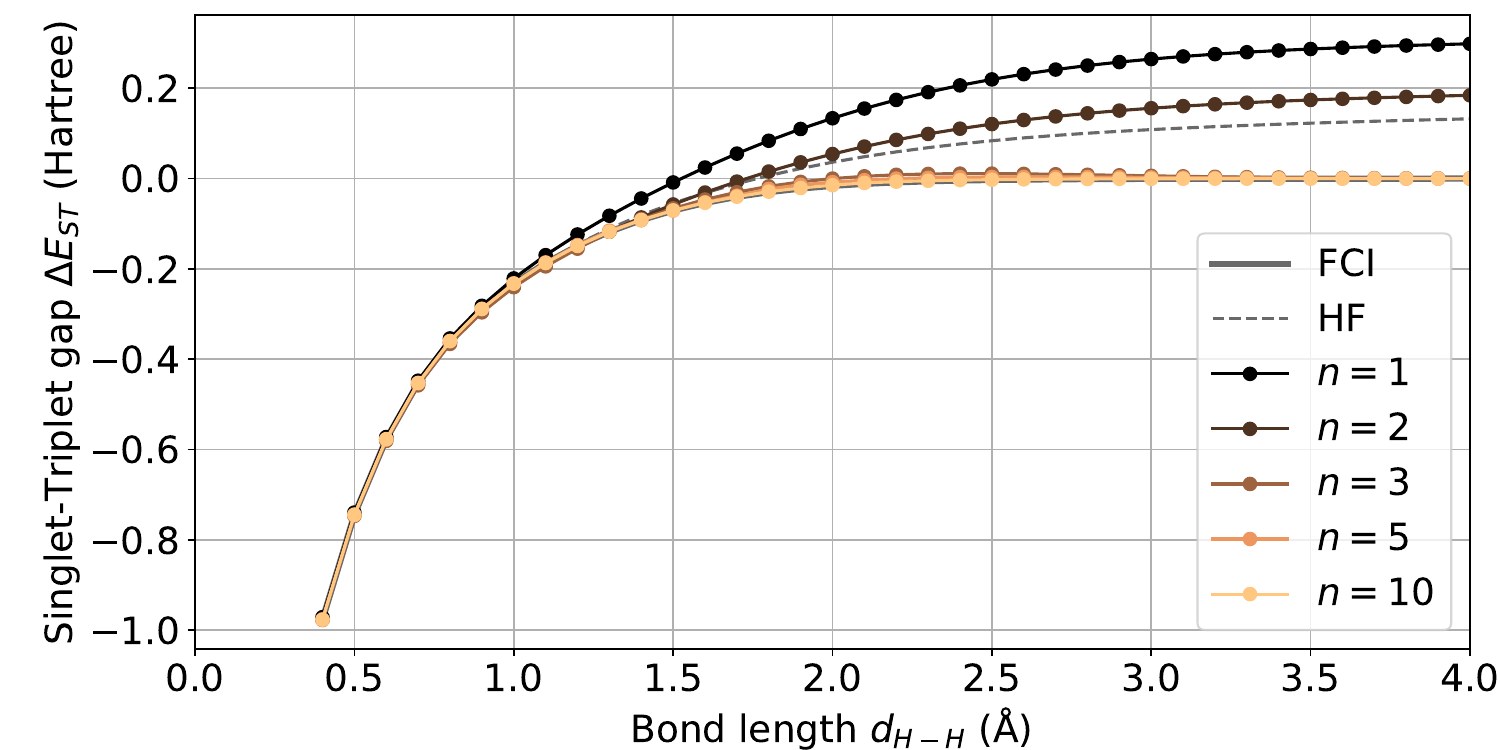}\\
 {\sffamily {\scriptsize (b)}}\includegraphics[width=0.99\columnwidth]{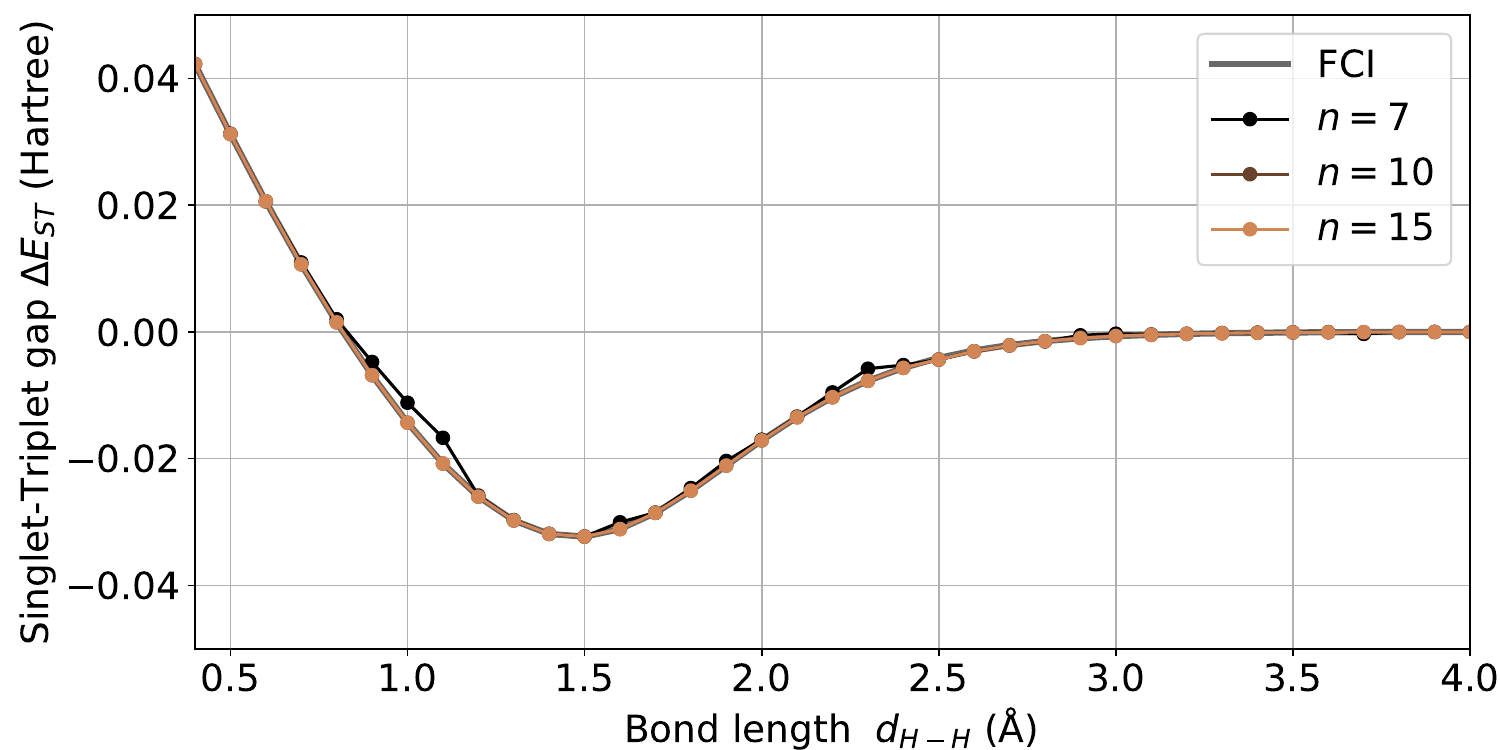}
  \caption{\textit{VQSM Singlet-Triplet Gap with $E_I$ Cost Function and a Single Layer HEA.}  
Singlet-Triplet gap $\Delta E_{ST} = E_T({\rm H}_4) + E_S({\rm H}_4)$ as a function of the interatomic distance $d_{\rm H-H}$ (in Å) at different iterations $n$ for the open (a) and closed (b) H$_4$ molecules. Results are compared with values obtained using the Hartree--Fock approximation and FCI method.
   }
  \label{fig:VQSM_STgap}
\end{figure}
Beyond the accurate estimation of ground-state energies, the presented IVQE and VQSM algorithm enables access to low-lying excited-state properties. In this section, we demonstrate the capability of VQSM to compute charge and spin gaps, which are essential descriptors of electronic excitation spectra and chemical reactivity. In the following we focus on VQSM using cost function $E_I(\bmtheta)$ with a single HEA layer, since this has been shown to converge rapidly without optimization issues.

Fig.~\ref{fig:VQSM_ChargeGap} shows the evaluation of the charge gap, defined as
\begin{equation}
\Delta E_c = E(\mathrm{H}_4^+) + E(\mathrm{H}_4^-) - 2E(\mathrm{H}_4),
\end{equation}
as a function of $d_{\mathrm{H}-\mathrm{H}}$. From a practical perspective, the energies of the charged states $E(\mathrm{H}_4^+)$ and $E(\mathrm{H}_4^-)$ are obtained independently from the ground-state energy of $E(\mathrm{H}_4)$ by constructing subspaces initialized from the corresponding charged Hartree--Fock states.

At low iteration numbers, the estimated charge gaps significantly deviate from the exact (FCI) results, especially in the dissociation regime ($d_{\rm H-H} > 2.5$~\AA), where electron correlation becomes dominant. 
This underestimation reflects the limited expressiveness of the Krylov subspace when only a small number of vectors is used. However, as the number of iterations increases, the method systematically improves, and the computed $\Delta E_c$ values converge towards the FCI results. 
Already at iteration $n=5$, the method captures the correct qualitative trend with respect to bond length, and by iteration $n=10$, the charge gap is in good quantitative agreement with the reference data across the entire bond-length range. 
In contrast, the Hartree--Fock approximation fails to capture the correct asymptotic behavior at large distances, converging to zero in the dissociation limit.

Let us now explore the singlet-triplet gap,
\begin{equation}
\Delta E_{ST} = E_T(\mathrm{H}_4) - E_S(\mathrm{H}_4),
\end{equation}
for both linear-chain and square geometries of the neutral $\mathrm{H}_4$ molecule. The gap is plotted in
Fig.~\ref{fig:VQSM_STgap} as a function of $d_{\rm H-H}$ for various iteration numbers ($n = 1$, 2, 3, 5, and 10 for the linear case, $n=7$, 10, and 15 for the square $H_4$ molecule), and is compared with both FCI and Hartree--Fock results.
In the linear geometry Fig.~\ref{fig:VQSM_STgap}(a), the Hartree--Fock approximation incorrectly predicts a singlet-triplet gap that increases from negative to positive values as the bond length increases, suggesting a ground-state spin crossover. 
However, FCI data confirm that the singlet remains the ground state throughout, with the spin gap decreasing to zero near dissociation due to diminishing electron-electron repulsion. 
VQSM shows rapid convergence with respect to the number of iteration, capturing the correct trend of $\Delta E_{ST}$ already at iteration $n=3$, and being nearly indistinguishable from the FCI reference over the entire bond-length range at iteration $n=10$.
For the square geometry Fig.~\ref{fig:VQSM_STgap}(b), the singlet-triplet gap remains small (on the order of millihartrees), and thus demands high numerical precision. 
Here too, VQSM exhibits excellent performance: at iteration $n=10$, the method reproduces both the curvature and the magnitude of the FCI reference. 
In this case, a triplet-to-singlet transition occurs near $d_{\mathrm{H}-\mathrm{H}} \sim 0.8$~\AA. 
In the $\mathrm{H}_4$ ring, all hydrogen atoms are closer together, enhancing the effects of Fock exchange, which stabilizes ferromagnetic interactions and thus a triplet configuration. The system thus undergoes a subtle competition between ferromagnetic exchange and electron delocalization across the ring, leading to a delicate spin gap highly sensitive to correlation effects.

\section{Conclusion}

In this manuscript, we have proposed alternative symmetry-preserving cost functions, $E_G({\bm\theta})$ and $E_I({\bm\theta})$, for the VQE algorithm, aiming to reduce optimization challenges, particularly when employing HEA. Since the ground-state energy is not directly minimized, ground-state properties can still be obtained by activating an iterative procedure. Two iterative algorithms have thus been derived: IVQE and VQSM. In both approaches, at each iteration $n$, a variational trial vector $|\widetilde{\Phi}_1^{(n)}({\bm\theta})\rangle$ is optimized by minimizing the considered cost function. The main distinction between the two algorithms lies in the additional classical numerical tasks required by VQSM, where the set $\left\{|\widetilde{\Phi}_1^{(p)}({\bm\theta})\rangle\right\}$, for $0 \leq p \leq n$, forms a reduced Krylov-like subspace. 

Both algorithms, along with the proposed cost functions $E_G({\bm\theta})$ and $E_I({\bm\theta})$, have been implemented and tested on the H$_4$ molecule, and compared against Full Configuration Interaction and Hartree--Fock reference results. Regarding the cost functions, both $E_G({\bm\theta})$ and $E_I({\bm\theta})$ perform similarly. However, $E_I({\bm\theta})$ has been favored throughout the manuscript due to its expected greater numerical stability and reduced sensitivity to optimization issues, arising from its linear properties. In particular, it has been shown that using HEA as an ersatz reflection can lead to a highly non-convex energy landscape, especially as the number of layers increases.

In terms of algorithmic performance, VQSM outperforms IVQE at the cost of modest additional classical computation, which remains tractable. Furthermore, a formal connection between VQSM and the classical Lanczos algorithm has been established, as both methods ideally yield the same tridiagonal representation of the Hamiltonian in the reduced subspace. Remarkably, even with a shallow, single-layer ansatz, the VQSM algorithm exhibits geometric convergence. In practice, a significantly deeper circuit would be required to fully match the performance of the Lanczos algorithm, potentially at the expense of optimization stability. As VQSM can be seen as a variational construction of the Krylov subspace, we expect that a small number of iterations is needed to achieve a given accuracy. Moreover, the algorithm has been shown to converge within a single iteration if the \textit{good-guess} condition \(\omega = 0.5\) is fulfilled. 

Nonetheless, a good trade-off can be achieved between (i) the number of iterations, (ii) the circuit depth, and (iii) the complexity of the optimization process, to reach a desired level of accuracy. This balance is particularly crucial in near-degenerate regimes, where enhancing circuit expressibility becomes necessary but computationally costly. In practical applications, choosing the minimal circuit depth that ensures sufficient convergence while preserving feasible optimization and coherence constraints on quantum hardware is essential for the efficient deployment of VQSM on near-term quantum devices.

Finally, we have highlighted the strength of the VQSM approach in accurately and efficiently capturing both charge and spin excitations using a compact variational ansatz. The method demonstrates systematic and robust convergence with iteration number, achieving chemical accuracy for both neutral and charged states using a limited number of vectors in the reduced subspace. These results position VQSM as a versatile variational quantum algorithm, extending beyond ground-state energy estimation to address spectroscopic features and electronic correlations in quantum many-body systems.

\section*{Acknowledgments}
The authors thanks D. Lacroix from IJCLab in Saclay (France) for fruitful discussions.
HAA is grateful to the Hubert Curien program (CAMPUS FRANCE) for the support as well ad the Abdus Salam ICTP for the  HPC facilities.  Computer time for this study was also provided by the computing facilities of the MCIA (Mésocentre de Calcul Intensif Aquitain). BS and MS acknowledge the support of the French Agence Nationale de la Recherche (ANR), under grant ANR-23-PETQ-0006.

\section*{Authors contribution}
Algorithm implementation and data acquisition: H.A. and M.S.(equal), A. P. (support); Concept and methodology: M.S. (lead) and B.S. (support); Supervision and project management: M.S.;  Writing original draft: M.S (lead) and B.S. (support). All authors edited and commented on the manuscript.

\appendix

\section{Hartree-Fock weight}
\label{app:HFweight}
\begin{figure}[h]
  \includegraphics[width=1\columnwidth]{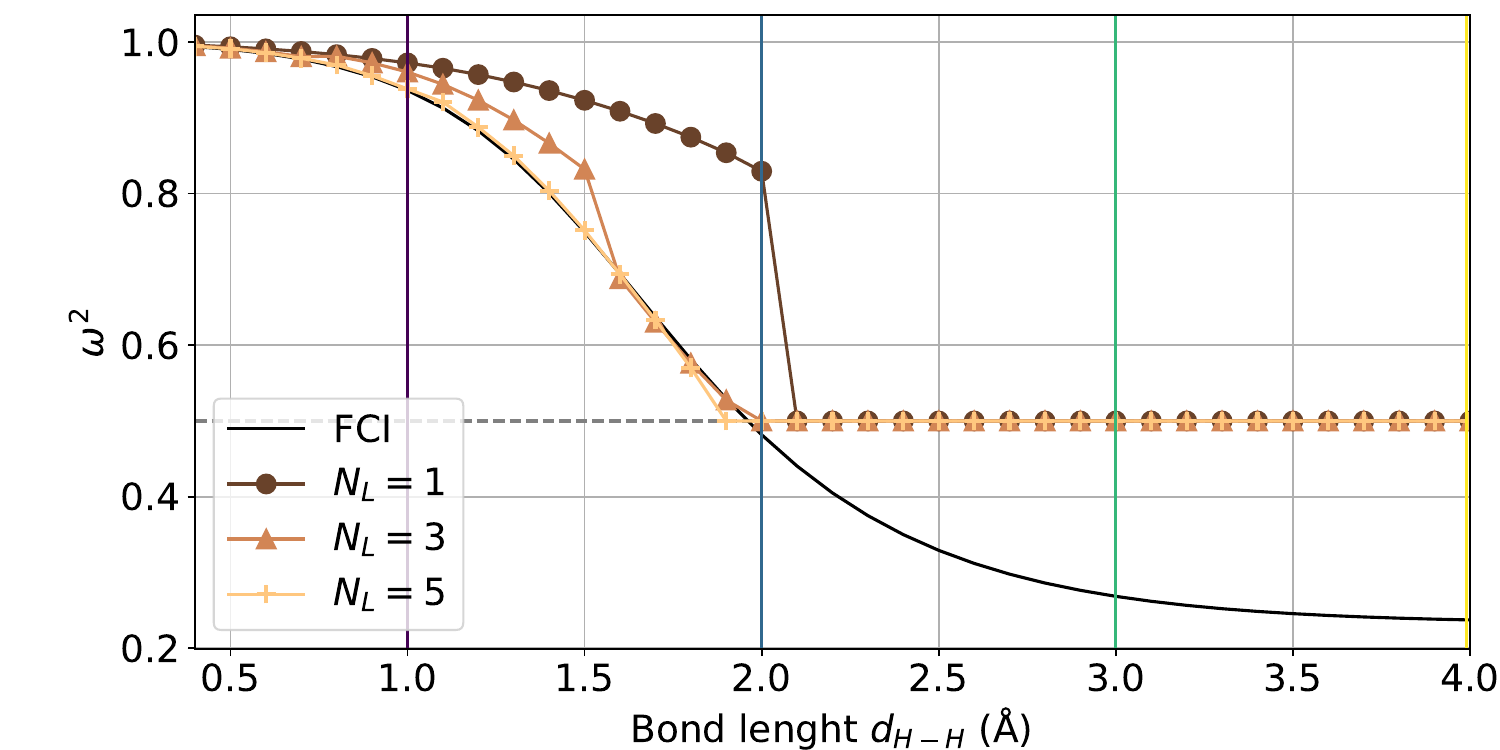}
  \caption{{\it Weight $w^2$  at the first iteration and as a function of the H-H bond length using the $E_G$ cost function}. Results are given considering different number of HEA layers $N_L$ and compared with the HF weight in the FCI ground state (black curve).}
  \label{fig:weights}
\end{figure}

Fig.~\ref{fig:weights} shows that $\omega^2(\bm\theta^*)$ saturates to a value of $1/2$ after only a few layers for $d_{\rm H-H} = 2$, $3$, and $4$~\AA. When $\omega^2 \leq 1/2$, the initial state $|\Phi_0\rangle$ is no longer a \textit{good state}, meaning it does not sufficiently overlap with the true ground state. As a result, the cost function $E_G(\bm\theta)$ fails to converge within a single iteration, necessitating the use of an iterative procedure.

%

\begin{thebibliography}{94}
\bibitem{fnote} An extension of that proof to non-Hermitian matrices is possible. Indeed, applying the Householder transformation $R$ to a non-Hermitian matrix $H$ leads to its lower Hessenberg form $\widetilde{H} = R H R$. Moreover, since $R$ is unitary, we have $\widetilde{H}^\dagger = R H^\dagger R$, which corresponds to the upper Hessenberg form of $H^\dagger$. Following the proof for Hermitian matrix, we can use that $[\widetilde{H}^\dagger \widetilde{H}]_{00} = [\widetilde{H}^\dagger(\boldsymbol{\theta}) \widetilde{H}(\boldsymbol{\theta})]_{00}$ such that  $ [\widetilde{H}^\dagger]_{01} [\widetilde{H}]_{10} \geq [\widetilde{H}^\dagger(\boldsymbol{\theta})]_{01} [\widetilde{H}(\boldsymbol{\theta})]_{10}$.
\end{thebibliography}

\end{document}